\documentclass[vecphys]{svmult}

\usepackage{makeidx}         % allows index generation
\usepackage{graphicx}        % standard LaTeX graphics tool
\usepackage{multicol}        % used for the two-column index
\usepackage{slashed}
\usepackage{amssymb}
\makeindex             % used for the subject index
\newcommand{\dsigmap}{\hat{\sigma}}
\def\wt#1{\widetilde{#1}}

\begin{document}

\title*{Two introductory lectures on high energy QCD and heavy ion collisions}
\author{Debasish Banerjee \inst{1}\and
 Jajati K. Nayak \inst{2}\and
 Raju Venugopalan \inst{3}}
\institute{Department of Theoretical Physics, Tata Institute of Fundamental Research, Homi Bhabha Road, Mumbai 400005, India
\texttt{debasish@theory.tifr.res.in}
\and Theoretical Physics Division, Variable Energy Cyclotron Centre, 1/AF Bidhan Nagar, Kolkata 700064, India  \texttt{jajati-quark@veccal.ernet.in}
\and Department of Physics, Bldg. 510A, Brookhaven National Laboratory, Upton NY 11973, USA. \texttt{raju@bnl.gov}}

\maketitle

\begin{abstract}
These introductory lectures present a broad overview of the physics of high parton densities in QCD and its application to our understanding of the early time dynamics in 
heavy ion collisions.

\end{abstract}

\section*{Introduction}
\label{sec:0}
Quantum Chromodynamics (QCD) is widely accepted as the fundamental theory describing the behavior of hadrons. In QCD, hadrons are composed of elementary quarks and gluons which are often together labeled partons. 
Quarks and gluons are never directly measured due to the confining property of the theory. However, their distributions inside a hadron can be probed precisely in Deep Inelastic Scattering (DIS) experiments. It is seen from the H1 and ZEUS experiments at HERA that the structure functions of the gluons and the sea quarks, which to leading order in QCD are related to their number densities, grow very rapidly \cite{zeus} with decreasing values of a
Lorentz invariant kinematic variable $x_{\rm Bj}$ introduced by Bjorken. Again, to lowest order, this variable corresponds to the momentum fraction x of the
hadron's momentum carried by a parton. Small x physics is the regime of high energies in QCD and the physics of this regime exhibits many
novel features that are not fully understood.

Small x physics is interesting for a variety of reasons. Even if the momentum transfers squared $Q^2$ are large enough such that, from asymptotic freedom, the QCD coupling constant $\alpha_s$ is small, the explosive growth in the number of partons with increasing energy makes the physics non-perturbative.~The strongest electric and magnetic fields found in nature occur in this situation.~The answer to a fundamental question in QCD -- can we calculate the total number of particles produced in a strong interaction at asymptotically high energies -- therefore requires that we understand the processes of particle creation in the presence of such strong color fields.
Particle production in this context is important for understanding a variety of striking but little understood phenomena at high energies. These
include, for instance,  i)limiting fragmentation \cite{limfrag},~where the rapidity distributions of the produced hadrons turns out to be independent of energy around the fragmentation region but possess non-trivial features in the central rapidity region, ii) the unusually large fraction of diffractive final states in DIS
where no particles are produced in angular regions relative to the scattering plane called ``rapidity gaps"~\cite{rapgap} and iii) the phenomenon of ``geometrical scaling",
where cross-sections appear to scale as a dimensionless function of the momentum transfer squared of the probe relative to a dimensionful dynamical
scale in the hadron~\cite{stasto}.  Small x physics is also of crucial importance in understanding the formation of bulk QCD matter in high energy heavy ion collisions and its possible thermalization to form a Quark Gluon Plasma (QGP). Finally, the hope is that understanding the properties of strong fields at small 
x will provide some insight into confinement in QCD and its role in high energy scattering. 

These questions can be addressed in a weak coupling effective field theory formalism~\cite{raju1,weigert} which describes the properties of high energy wavefunctions as a Color Glass Condensate (CGC). In this approach, the degrees of freedom in the high energy nuclear wavefunction are
divided into static light cone color sources at large x and dynamical gauge fields at small x which are coupled to these static color sources.
Because the scale between the two sorts of degrees of freedom is arbitrary, and because physics cannot depend on this scale, one obtains a  renormalisation group description in rapidity which arises from successively integrating out dynamical fields at one scale and absorbing them into color
sources at the next. 

Efforts to understand the rich non-perturbative phenomena of high energy QCD are the subject of these lectures.~The first lecture begins with
DIS and describes high energy scattering in DIS in the well understood Bjorken limit and the much less understood Regge-Gribov limit of QCD.~We show
that parton saturation arises naturally in the latter limit.~The effective field theory formalism of CGC, which incorporates the physics of parton saturation,
is described next. We then discuss color dipole models that incorporate simply both the non-linear dynamics of saturation and the linear dynamics of
perturbative QCD in DIS and hadron scattering.~In the second lecture,~we apply the CGC approach to treat {\it ab initio} heavy ion collisions at high energies. When two sheets of colored glass collide, they form bulk QCD matter called the  Glasma in heavy ion collisions. We describe the properties of the Glasma, probes of its dynamics and the possible fast thermalization of the Glasma into a Quark Gluon Plasma.
%%%%%%%%%%%%%%%%%%%%%%%%%%%%%%%%%%%%%%%%%%%%%%%%%%%%%%%%%%%%%%%%

\section{Deep Inelastic Scattering}
\begin{figure}
\centering
\begin{tabular}{cc}
  \resizebox{50mm}{!}{\includegraphics{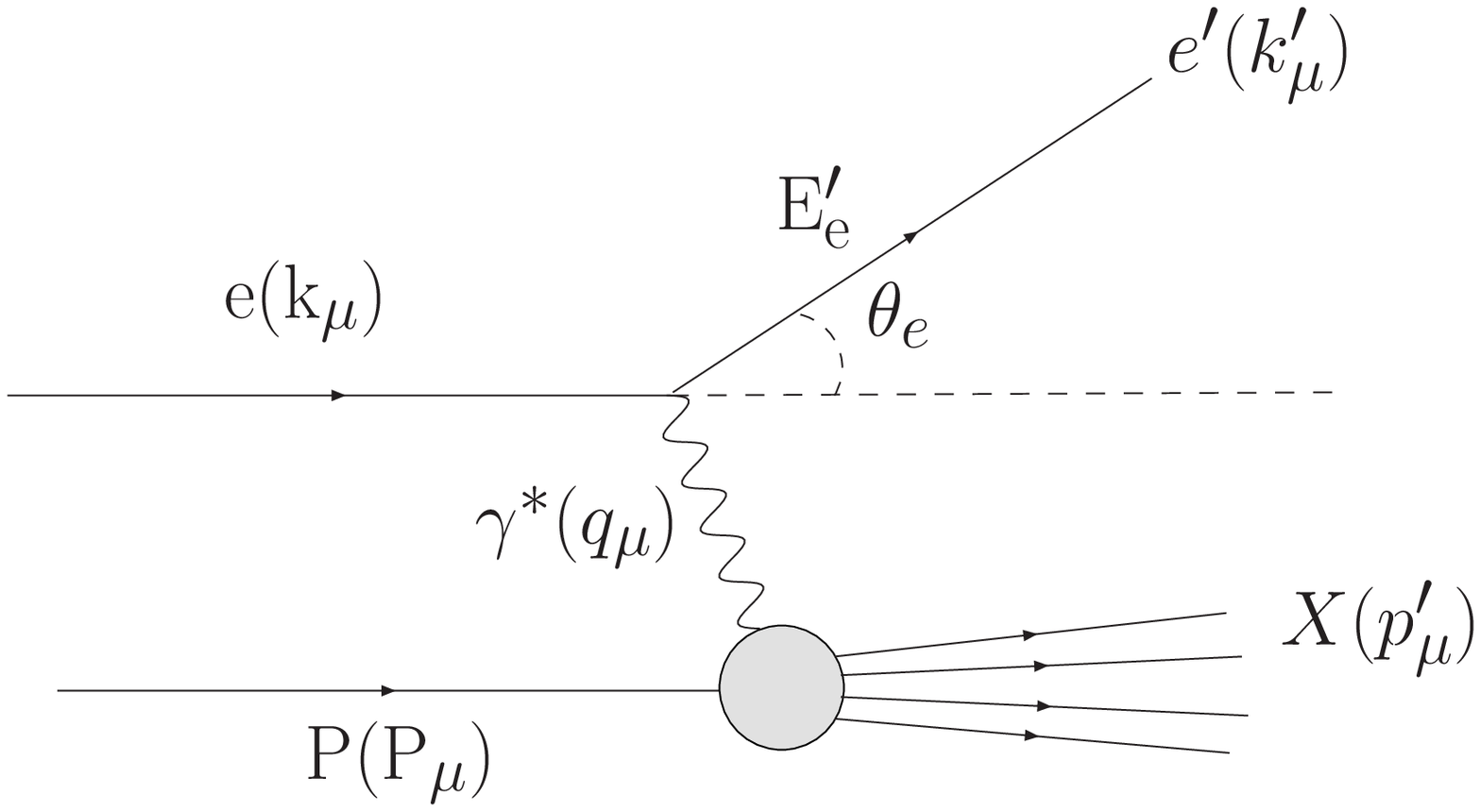}}&
  \resizebox{50mm}{!}{\includegraphics{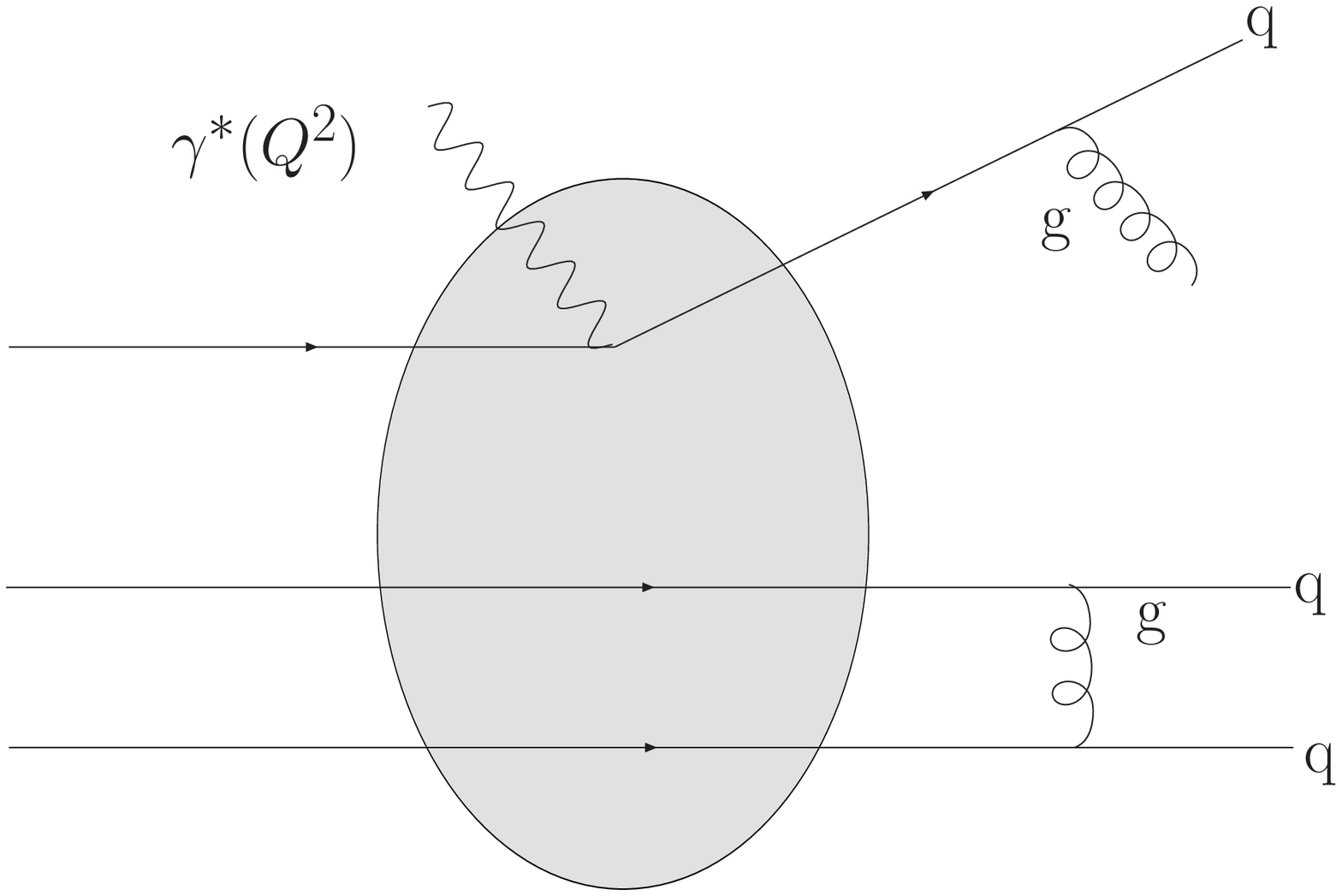}}
 \end{tabular}
\caption{~Left.~The general setup of DIS.~Right.~Inside the hadron via the Parton model.~In the impulse approximation the partons inside are non-interacting on the interaction time-scale of the virtual photon probe.}
\label{fig:feynd1}       % Give a unique label
\end{figure}

Deep Inelastic Scattering (DIS)  involves the scattering of a high energy lepton off a hadronic target in which the energy and momentum transfer of the lepton is measured experimentally.~DIS is essentially a two step process where the lepton emits a virtual photon in the first step which then interacts with the hadron in the second step.~Depending on the energy transferred,~the proton can break up and new particles are created.~This provides a clean environment to study the structure of hadrons at high energies since one of the interaction vertices in the two step process is completely known.
The DIS process is schematically illustrated in fig.~\ref{fig:feynd1}.~The cross section of the process is
%%%%%%%%%%%%%%%%%%%%%%%%%%%%%
\begin{equation}
 \frac{d^2\sigma^{eh \rightarrow eX}}{dx dQ^2}=\frac{4\pi \alpha^2_{em}}{x Q^4}\left[(1-y+\frac{y^2}{2})F_2(x,Q^2)-\frac{y^2}{2}F_L(x,Q^2)\right]\, .
\end{equation}
%%%%%%%%%%%%%%%%%%%%%%%%%%%%%

%%%%%%%%%%%%%%%%%%%%%%%%%%%%%%
\framebox[11.0cm]{ \parbox[t]{10cm}{ {\bf Light Cone coordinates :} It is useful at this point to introduce light cone (LC) coordinates which are very useful to discuss high energy scattering.~Let z denote the longitudinal axis of collision.~For an arbitrary 4-vector $a^{\mu}=(a^t,a^x,a^y,a^z)$,~the LC coordinates are defined as:
%%%%%%%%%%%%%%%%%%%%%%%%%%%%%%
\begin{equation}
 a^+ \equiv \frac{a^t+a^z}{\sqrt{2}},a^- \equiv \frac{a^t-a^z}{\sqrt{2}},a_{\perp} \equiv (a^x,a^y).
\end{equation}
%%%%%%%%%%%%%%%%%%%%%%%%%%%%%%
In particular $x^+=(t+z)/\sqrt{2}$ is the LC ``time'' and $x^-=(t-z)/\sqrt{2}$ is known as the LC ``longitudinal coordinate''.
}}
\vskip 1.0cm
Here $e$ and $h$ denote the initial electron and hadron state and x the final hadronic state.~$\alpha_{em}$ is the usual QED coupling constant.~
The four vectors of the electron and proton are shown in fig.~\ref{fig:feynd1}. If $q^{\mu}$ denotes the (space-like) 4-momenta of the exchanged photon,~then the energy transferred to the hadron is
%%%%%%%%%%%%%%%%%%%%%%%%%%%%%%
\begin{equation}
 Q^2=-q^{\mu} q_{\mu}=-(k_{\mu}-k_{\mu}^{\prime})^2=4 E_e E_e^{\prime} \sin^2\left(\frac{\theta^{\prime}_e}{2}\right)\, .
\end{equation}
%%%%%%%%%%%%%%%%%%%%%%%%%%%%%%
The $Q^2$ is a measure of the resolution power of the probe.~$E_e$ and  $E_e'$ are the initial and the final energies of the lepton and $\theta_e '$ is the lepton scattering angle in the center-of-mass(COM) frame.~The variable $y$ provides a measure of the inelasticity of the collision and is defined in the following frame invariant way
%%%%%%%%%%%%%%%%%%%%%%%%%%%%%%
\begin{equation}
 y=\frac{P.q}{P.k}=1-\frac{E_e^{\prime}}{E_e}\cos^2\frac{\theta_e^{\prime}}{2}
\end{equation}
%%%%%%%%%%%%%%%%%%%%%%%%%%%%%
The x here is the Bjorken variable $x_{\rm Bj}$ whose frame invariant definition is
%%%%%%%%%%%%%%%%%%%%%%%%%%%%%
\begin{equation}
 x_{Bj}=\frac{Q^2}{2P.q}
\end{equation}
%%%%%%%%%%%%%%%%%%%%%%%%%%%%%
At very high energies, we obtain $x_{\rm Bj} \sim Q^2/s$ for a fixed $y$.~This clarifies why, for a fixed $Q^2$,  small x physics is the physics of high
energies in QCD. As we shall see shortly, $x_{\rm Bj}$ is related to the momentum fraction of the struck quark.

\begin{figure}
\centering
\includegraphics[scale=0.5]{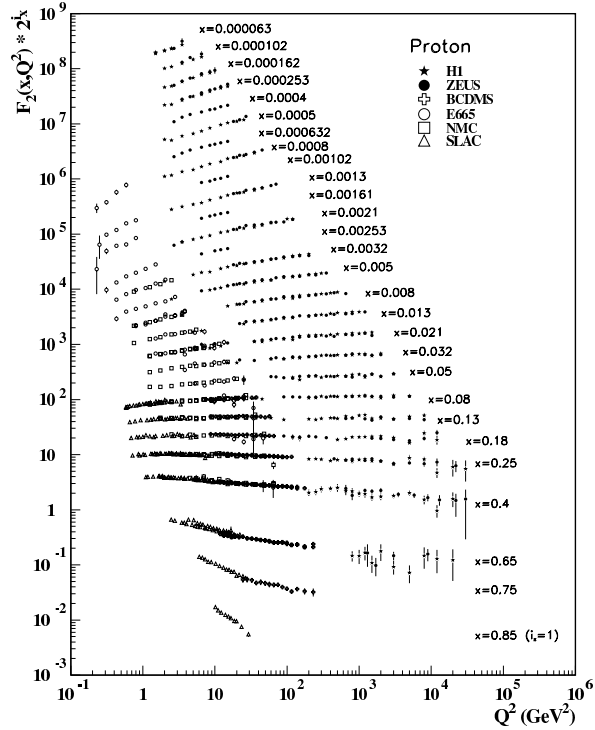}
\caption{Bjorken Scaling:~Independence of structure functions vs $Q^2$ in proton.~Courtesy PDG.}
\label{fig:Bj-DIS}       % Give a unique label
\end{figure}

$F_2$ is the structure function that, at leading order in QCD, gives the quark $+$ antiquark distributions in a proton,~while the longitudinal structure function $F_L$ is a measure of the gluon momentum distribution.~These quantities can be independently extracted from the DIS cross section by varying
x, $Q^2$ and the center of mass energy of the collision.~The most striking feature observed in early DIS experiments was the flatness of the structure function $F_2$ over a wide range of $Q^2$ at which the experiments were first performed (see figure \ref{fig:Bj-DIS}).~The apparent scale invariance of the structure function gave the needed experimental support for the hypothesis of point-like, weakly interacting partons inside hadrons.~The dynamical model that was formulated by Feynman \cite{feyn} to understand this phenomena is called the parton model.
%%%%%%%%%%%%%%%%%%%%%%%%%%%%%
%%%%%%%%%%%%%%%%%%%%%%%%%%%%%

\subsection{The Bjorken limit, the Parton model and pQCD}
The Bjorken limit, in which the results of the DIS experiments were correctly interpreted, is attained when the center of mass energy $s \rightarrow \infty$ and the energy transferred in the collision $Q^2 \rightarrow \infty$ keeping the ratio $x_{\rm Bj}=Q^2/s={\rm fixed}$.~At these very high energies,~the rest mass energy of the hadronic target $M$ can be neglected compared to its longitudinal momentum $P^+$.~In the parton model, formulated in the ``infinite-momentum'' frame, the hadron can be thought of as a collection of ``quasi-free'' partons which are nearly on-shell excitations carrying some fraction $x_F$ of the total hadron momentum.~Moreover the entire momentum of the hadron can be thought to be longitudinal.~This picture is essentially true since the interactions between the partons are highly time-dilated.~In the ``impulse approximation'' when the virtual photon strikes a parton,~the other partons form the spectators without interacting with the struck quark or among themselves. (See fig.~\ref{fig:feynd1}).~This demands
%%%%%%%%%%%%%%%%%%%%%%%%%%%%%%%%
\begin{equation}
 (x_F P+q)^2=m_q^2 \simeq 0;\,\,\,\, x_F \simeq \frac{Q^2}{2P.q}\equiv x_{\rm Bj}\, ,
\end{equation}
%%%%%%%%%%%%%%%%%%%%%%%%%%%%%%%%
which confirms our interpretation of $x_{\rm Bj}$.

The naive parton model also predicts $F_L$ to be zero. ~This result is called the Callan-Gross relation and it provides strong evidence that the partons
probed by the virtual photon are spin-$1/2$ objects.~It was also realized that the hadron might also contain an infinite \emph{sea} of light $q \overline{q}$ pairs,~called appropriately \emph{sea quarks}, as opposed to the \emph{valence quarks} that carry the net baryon number of a hadron.~However it was experimentally found that, at a scale of O(1 GeV), the proposed valence and sea quark distributions could only account for about $50 \%$ of the total momentum in a proton.~This necessitated the existence of other partons (\emph{gluons}) which in turn explained the puzzle that $F_L$ was not zero experimentally.~With further experiments, it was also demonstrated that the Bjorken scaling was only approximately true and there were scaling violations at lower-x.~This did not have an explanation within the naive parton model at all.

\begin{figure}
\centering
\includegraphics[scale=0.35]{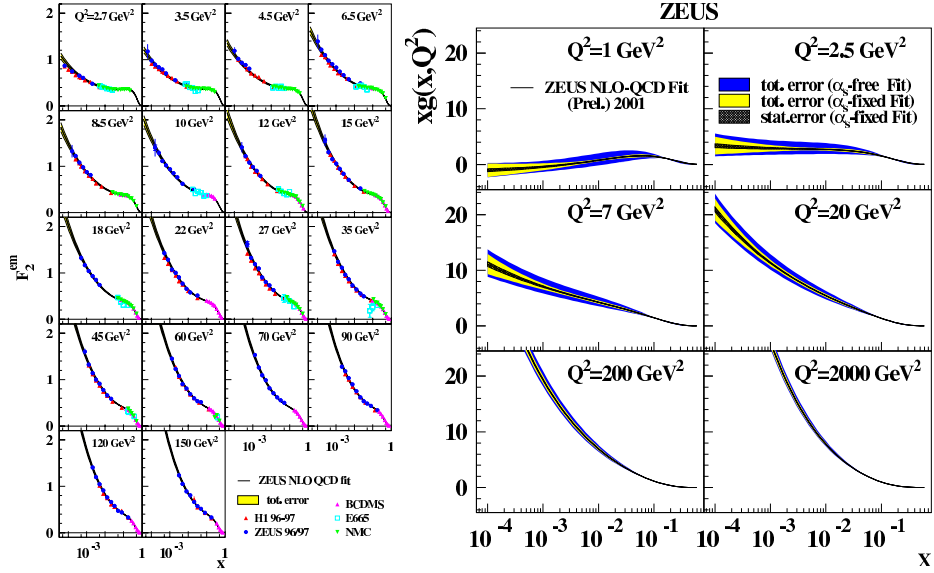}
\caption{Rapid growth of the structure functions at low x from the ZEUS experiment at HERA.}
\label{fig:dglap1}       % Give a unique label
\end{figure}

These experimental results set the stage for the formulation of the theory of strong interactions in terms of QCD which had quarks and gluons. Most important though, in tying all the pieces together, was the (Nobel) discovery of Gross, Wilczek and Politzer that the theory had a $\beta$ function with negative sign-- thereby indicating that the coupling asymptotically goes to zero at high $Q^2$. Perturbative QCD (pQCD) therefore showed that in the Bjorken limit the naive parton model is indeed a good approximation.~It went further to give a quantitative explanation of the log scaling violations that were observed in the experiments.~The key idea to the solution of this puzzle lay in the assumption of neglecting the transverse momentum of a quark.~In fact,~a quark can emit a gluon and acquire a large transverse momentum $p_T$, on time scales comparable with that of the probe, with a probability proportional to $\alpha_s dp_T^2/p_T^2$ at large $p_T$ where $\alpha_s$ is the strong coupling constant.~On integrating this till the kinematic limit of $p_T^2 \sim Q^2$,~contributions proportional to $\alpha_s \ln Q^2$ are obtained which are precisely the scaling violations.

pQCD now has a host of machinery dedicated to precision physics.~Tools such as the Operator Product Expansion~(OPE) are used to calculate many observables upto high orders in perturbation theory.~Factorization theorems have been derived to separate out soft and hard dynamics in a systematic way; the former are parameterized as non-perturbative parton distribution functions while the latter can be computed order by order in perturbation theory.
The factorization of hard and soft scales is manifest in a renormalisation group treatment that requires that physics be independent of this scale. An important 
result is the  Dokshitzer-Gribov-Lipatov-Altarelli-Parisi (DGLAP)~\cite{dglap} equation for the evolution of quark and gluon distribution functions as a function of $Q^2$ which made earlier work on the OPE accessible to straightforward experimental analysis. 
~A consequence of DGLAP evolution is that, as shown in fig.~\ref{fig:dglap1}, the parton distributions  increase rapidly at small-x as $Q^2$ increases.~This means that as one probes finer and finer transverse resolution scales, more and more partons~(which share the total hadron momentum) can be resolved within the hadron.~However, albeit the number of partons increases with increasing $Q^2$, the phase space density--the number of partons/area/$Q^2$--decreases and the proton becomes more and more dilute.
%%%%%%%%%%%%%%%%%%%%%%%%%%%%%%
%%%%%%%%%%%%%%%%%%%%%%%%%%%%%%%
\subsection{The Regge-Gribov limit in QCD and Saturation}

The DGLAP framework in pQCD has worked very well and explains many features of the HERA data very well. However, when pushed to
lower $Q^2$, one begins to note increasingly unpleasant results. As shown in the right plot of fig.~\ref{fig:dglap1}, fits to data push the gluon distributions
into negative territory. This also seems to afflict $F_L$, which much be a positive definite quantity as it is directly proportional to a cross-section.
Another feature of HERA data that the conventional pQCD approach needs further parameters to explain is the high fraction of diffractive events and the fact that these diffractive events have the same dependence as the total cross-section on the energy of the photon-hadron system. Within pQCD itself, it is expected that DGLAP should be supplemented by power corrections in $Q^2$ called "higher twist" effects, which become increasingly important 
as $Q^2$ is decreased and as one goes to smaller values of x. The higher twist formalism is however very cumbersome and not under theoretical control. It is therefore useful to take stock of the physics underlying the dynamics of partons in this kinematic region and approach the problem anew. 

The physics of the small x regime is best understood in a very different asymptotic limit from the better known (and understood) Bjorken limit and goes by the name of Regge-Gribov limit in QCD.~This limit corresponds to ~$x_{\rm Bj}\rightarrow0$ and $s\rightarrow \infty$ with $Q^2(\gg \Lambda_{QCD}^2)={\rm fixed}$.~The Regge-Gribov limit corresponds to the regime of strong color fields in QCD. It is responsible for the bulk of multi-particle production in QCD. At sufficiently large $Q^2 \gg \Lambda_{\rm QCD}^2$, the physics of this regime, albeit non--perturbative, should be accessible in weak coupling.

\begin{figure}
\centering
\includegraphics[scale=0.7]{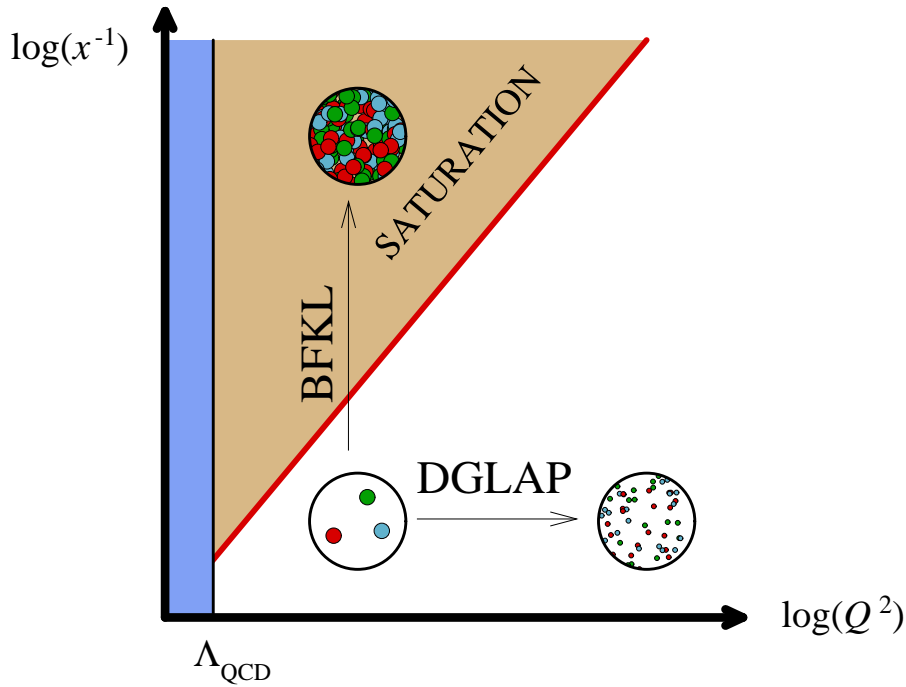}
\caption{The horizontal and vertical directions correspond respectively to the DGLAP and BFKL regimes of QCD. The diagonal line corresponds to the
saturation boundary. The thin vertical region to the left is the non-perturbative region.}
\label{fig:dglap+bfkl}       % Give a unique label
\end{figure}

In the Regge--Gribov limit, the leading logarithms in x, $\alpha_S \ln(1/x)$ dominate over the leading logs $\alpha_S \ln(Q^2/\Lambda_{\rm QCD}^2)$--the converse is true of course in the Bjorken limit. The renormalization group equation that resum these leading logs of x  is called
the Balitsky--Fadin--Kuraev--Lipatov (BFKL) equation~\cite{bfkl}. The BFKL equation leads to a rapid power law growth in the gluon distribution with
decreasing x. A schematic diagram of the parton content of the hadron in the two asymptotic limits is shown in fig.~\ref{fig:dglap+bfkl}.
%%%%%%%%%%%%%%%%%%%%%%%%%%%%%%%%%%
%%%%%%%%%%%%%%%%%%%%%%%%%%%%%%%%%%%
The physics issues at small-x have a simple intuition.~Both theory and experiments show that the gluon and the sea quark densities grow very rapidly in low-x.~As parton distributions grow, for fixed $Q^2$, the occupation number of gluons in
the hadron wavefunction becomes increasingly large and it is no longer possible to neglect their mutual interactions.~In terms of Feynman diagrams,~ these
interactions involve gluon recombination and screening which deplete the gluon density relative to the bremsstrahlung diagrams that are responsible for the
rapid growth of the parton distributions.~The net effect of the competition between these two effects is that the occupation number of partons ``saturates''
at the maximum possible value of $1/\alpha_s$ in QCD. The dynamics of QCD in this regime is fully non--linear corresponding to the strongest electric and magnetic fields in nature.~Each mode in the wavefunction, because it has a different occupation number, will saturate at a different value of x--the
momentum scale $Q_s$ at which it saturates is called the saturation scale~\cite{glr}. In terms of the gluon distribution, the maximal phase
space density is reached when
%%%%%%%%%%%%%%%%%%%%%%%%%%%%%%%%%
\begin{equation}
 {1\over 2(N_c^2 -1)}\, {x G(x,Q_s^2)\over \pi R^2 Q_s^2} = {1\over \alpha_S(Q_s^2)} \, ,
\label{eq:sat1}
\end{equation}
%%%%%%%%%%%%%%%%%%%%%%%%%%%%%%%%%%
where $\pi R^2$ is the hadron area in the impact parameter space which is only well defined if the wavelength of the probe is small compared to $R$,~which we will assume throughout in the future discussions. As suggested by eq.~\ref{eq:sat1}, the growth of the gluon distribution functions with decreasing x implies a growth of the saturation scale as well. The saturation scale is the typical momentum of gluons in the high energy hadron wavefunction, and
when x is small, $Q_s^2 \gg \Lambda_{\rm QCD}^2$ is a semi-hard scale in QCD. Thus $\alpha_s(Q_s^2)<<1$,  suggesting that  weak coupling
techniques can be applied to study the saturation regime in the Regge--Gribov limit of QCD.~The use of weak coupling techniques does not necessarily mean that the physics is perturbative; in this case,~the many-particle interactions are responsible for the non-perturbative nature of saturation, because, although individual interactions are small,~the large number of gluons amplifies the effect necessitating a resummation of high parton density effects.

To summarize, saturation is a natural consequence of QCD in the Regge-Gribov limit of the theory because gluon occupation numbers cannot be
arbitrarily large in the theory. Because saturation is achieved for different modes at different values of x, this also naturally suggests a dynamical scale
that characterizes the onset of the non-linear saturation dynamics in QCD. If this scale is large compared to the QCD confining scale $\sim \Lambda_{\rm QCD}$, the physics of saturation can be described using a weak coupling formalism. Because occupation numbers are large in this regime, the appropriate
degrees of freedom here are classical fields. One is therefore, on very general grounds, led to postulate the existence of a weak coupling effective theory
that captures the physics of this intensely non-linear regime of QCD. This is the Color Glass Condensate (CGC), and as we shall discuss, its name
captures the properties of the matter in the hadron wavefunctions at high energies.

%%%%%%%%%%%%%%%%%%%%%%%%%%%%%%%%%%%%%%
\subsection{The Color Glass Condensate}
One way to approach the small x problem is to appreciate that there is a formal Born-Oppenheimer separation between large x and small x
modes~\cite{Susskind} for quantum field theories on the light cone. These are respectively the slow and fast modes in the effective theory. Thus on the time scale of the ``wee" parton small x fields, the large x partons can be viewed as static charges.
Since these are color charges, they cannot be integrated out of the theory but must be viewed as
sources of color charge for the dynamical wee fields. With this dynamical principle in mind, one
can write down an effective action for wee partons in QCD at high energies~\cite{MV}. The generating functional
of wee partons has the form
\begin{eqnarray}
{\cal Z}[j] = \int [d\rho]\, W_{\Lambda^+}[\rho]\,\left\{ {\int^{\Lambda^+} [dA] \delta(A^+) e^{iS[A,\rho]-j\cdot A} \over \int^{\Lambda^+} [dA] \delta(A^+) e^{iS[A,\rho]}}\right\}
\label{eq:genfunc}
\end{eqnarray}
where the wee parton action has the form
\begin{eqnarray}
S[A,\rho] ={-1\over 4}\,\int d^4 x \,F_{\mu\nu}^2
+ {i\over N_c}\, \int d^2 x_\perp dx^- \delta(x^-)
\times {\rm Tr}\left(\rho(x_\perp)U_{-\infty,\infty}[A^-]\right) \, .
\label{eq:action}
\end{eqnarray}
In Eq.~\ref{eq:genfunc}, $\rho$ is a classical color charge density  of the static sources and
$W[\rho]$ is a weight functional of sources (which sit at momenta $k^+ > \Lambda^+$: note,
$x = k^+/P_{\rm hadron}^+$). The sources are coupled to the dynamical wee gluon fields (which in
turn sit at $k^+ < \Lambda^+$) via the gauge invariant term\footnote{An alternative gauge invariant form of the coupling of sources and fields is obtained in Ref.~\cite{JJV}-it  reproduces the BFKL equation more efficiently.} which is the second term on the RHS of
Eq.~\ref{eq:action}. The first term in Eq.~\ref{eq:action} is the QCD field strength tensor squared-- the wee
gluons are treated in full generality in this effective theory, formulated in the light cone gauge $A^+=0$. The source $j$ is an external source-derivatives taken with respect to this source (with this source then put to zero) generate correlation functions in the effective
theory.

The argument for why the sources are classical is subtle and follows from a coarse graining of the effective action to only include modes of interest. For large nuclei, or at small x, wee partons couple to a large number of sources. For a large nucleus, it can be shown
explicitly that this source density is classical~\cite{SR}. Further, it was conjectured that the weight functional for a large nucleus was a Gaussian in the source density (corresponding to the quadratic Casimir operator)~\cite{MV,Kovchegov}. This was shown explicitly later to be correct--albeit with a small but interesting correction, proportional to the cubic Casimir operator, that  generates Odderon excitations in the effective theory~\cite{SR}.
For a large nucleus, the variance of the Gaussian distribution, the color charge squared per unit area $\mu_A^2$,
proportional to $A^{1/3}$, is a large scale-and is the only scale in the effective action\footnote{$\mu_A$ is simply related to $Q_s$: $Q_s \sim 0.6\,\mu_A$.
For a detailed discussion, see Ref.~\cite{Tuomas}.}. Thus for $Q_s^2 >> \Lambda_{\rm QCD}^2$,
$\alpha_S(\mu_A^2) <<1$, and one can compute the properties of the theory in Eq.~\ref{eq:genfunc} in weak coupling.

The Yang-Mills equations can be solved analytically to obtain the classical field of the nucleus as a function of $\rho$: $A_{\rm cl.}(\rho)$~\cite{MV}. From the generating functional in Eq.~\ref{eq:genfunc}, one obtains for the two point correlator,
\begin{eqnarray}
<A A> = \int [d\rho]\,W_{\Lambda^+}[\rho]\, A_{\rm cl.}(\rho)\,A_{\rm cl.}(\rho) \, .
\label{eq:YM1}
\end{eqnarray}
From this expression, one can determine (for Gaussian sources) the occupation number $\phi = dN/\pi R^2/ dk_\perp^2 dy$ of wee partons in the classical field of the nucleus.
For $k_\perp >> Q_s^2$, one has the Weizs\"acker-Williams spectrum $\phi \sim Q_s^2 / k_\perp^2$, while for $k_\perp \leq Q_s$, one has a  resummation to all
orders in $k_\perp$, which gives $\phi\sim {1\over \alpha_S} \ln(Q_s/k_\perp)$. (The behavior at low $k_\perp$ can, more accurately, be represented
as ${1\over \alpha_S} \Gamma(0,z)$ where $\Gamma$ is the incomplete Gamma function and $z = k_\perp^2/Q_s^2$.) A  nice expression for the classical field of the nucleus containing these two limits is given in Ref. ~\cite{Dionysis}.

We are now in a position to discuss why a high energy hadron behaves like a Color Glass Condensate~\cite{raju1}. The ''color''  is obvious since the degrees of freedom, the partons, are colored. It is a glass because the stochastic sources (frozen on time scales much larger than the wee
parton time scales) induce a stochastic (space-time dependent) coupling between the partons under quantum evolution (to be discussed in the next section)-this is analogous to a spin glass in condensed matter physics. Finally, the matter is a condensate since the wee partons have large occupation numbers (of order $1/\alpha_S$) and have momenta peaked about $Q_s$. Just as in actual condensates, the number of gluons, for a fixed configuration of sources,
has a non--zero value and has a magnitude of order $1/\alpha_S$. Gauge invariant observables are computed by averaging over all possible configurations.
As we will discuss, these properties are enhanced by quantum evolution in x. The classical field retains its structure-while the saturation scale grows: $Q_s(x^\prime) > Q_s(x)$ for $x^\prime < x$.

The problem of small fluctuations about the effective action in Eq.~\ref{eq:action} were first addressed in ref.~\cite{AJMV} and it was noted that these 
gave large corrections of order $\alpha_S\ln(1/x)$ to the classical action; this implies that the Gaussian weight functional is fragile under quantum evolution of the sources. A Wilsonian renormalization group (RG) approach was later developed that systematically
treated these corrections~\cite{JIMWLK}. The basic recipe is as follows. Begin with the generating functional in
eq.~\ref{eq:genfunc} at some $\Lambda^+$, with an initial source distribution $W[\rho]$. Perform small fluctuations about the classical saddle
point of the effective action, integrating out momentum modes in the region ${\Lambda^\prime}^+ < k^+ < \Lambda^+$,  ensuring that
${\Lambda^\prime}^+$ is such that $\alpha_S \ln(\Lambda^+/{\Lambda^\prime}^+) << 1$. The action reproduces itself at the new
scale ${\Lambda^\prime}^+$, albeit with a charge density $\rho^\prime = \rho + \delta \rho$,  where $\delta \rho$ has support only in the window ${\Lambda^\prime}^+ < k^+ < \Lambda^+$, and $W_{\Lambda^+}
[\rho]\longrightarrow W_{{\Lambda^\prime}^+}[\rho^\prime]$. The change of the weight functional $W[\rho]$ with x is described by
the JIMWLK- non-linear RG equation~\cite{JIMWLK} which we shall not  write explicitly here--it can be found for instance in ref.~\cite{raju1,weigert}. The 
JIMWLK equation has been re--derived subsequently by several authors. We will discuss briefly in the next lecture, one such derivation by one of us and 
collaborators, in the  context of nucleus--nucleus collisions.

The JIMWLK equations form an infinite hierarchy (analogous to the BBGKY hierarchy in statistical mechanics) of
ordinary differential equations for the gluon correlators $<A_1 A_2 \cdots A_n>_Y$, where $Y= \ln(1/x)$ is the rapidity. The expectation
value of such an operator ${\cal O}$ is defined to be
\begin{eqnarray}
\langle{\cal O}\rangle_Y = \int [d\alpha] {\cal O}[\alpha] W_Y[\alpha] \,,
\label{eq:4}
\end{eqnarray}
where $\alpha = {1\over \nabla_\perp^2}\,\rho$. The corresponding JIMWLK equation for this operator is
\begin{eqnarray}
{\partial \langle{\cal O}[\alpha]\rangle_Y\over \partial Y} = \langle{1\over 2}\,\int_{x_\perp,y_\perp}\,{\delta \over \delta \alpha_Y^a(x_\perp)}\,\chi_{x_\perp,y_\perp}^{ab}[\alpha] {\delta \over \delta \alpha_Y^b (y_\perp)} {\cal O}[\alpha]\rangle_Y \,.
\label{eq:5}
\end{eqnarray}
Here $\chi$ here is a non-local object expressed in terms of path ordered (in rapidity) Wilson lines of $\alpha$~\cite{raju1}. This equation is analogous to a
(generalized) functional  Fokker-Planck equation, where $Y$ is the "time" and $\chi$ is a generalized diffusion coefficient. This equation illustrates
the stochastic properties of operators in the space of gauge fields at high energies. For the gluon density, which is proportional to a two-point function
$<\alpha^a (x_\perp) \alpha^b (y_\perp)>$, one recovers the BFKL equation in the limit of low parton densities. 
For a first attempt to solve the JIMWLK equation numerically, see ref.~\cite{RW}.

For large $N_c$ and large A ($\alpha_S^2 A^{1/3} >> 1$),  the expectation value of the product of traces of Wilson lines factorizes into the product of the expectation values of the traces:
\begin{equation}
<{\rm Tr}(V_x V_z^\dagger) {\rm Tr}(V_z V_y^\dagger)> \longrightarrow <{\rm Tr}(V_x V_z^\dagger)>\,<{\rm Tr}(V_z V_y^\dagger)> \, ,
\label{eq:6}
\end{equation}
where $V_x={\cal P}\exp\left(\int dz^- \alpha^a(z^-,x_\perp) T^a\right)$. Here ${\cal P}$ denotes path ordering in $x^-$ and $T^a$ is the SU(3) generator in the adjoint representation. In  the dipole picture, the cross-section for a dipole scattering off a target $P$ can be expressed in terms of these 2-point dipole operators as~\cite{Mueller1989:st}
\begin{equation}
\sigma_{q\bar q P} (x, r_\perp) = 2\, \int d^2 b\, \, {\cal N}_Y (x,r_\perp,b) \, ,
\label{eq:sigmaqqP}
\end{equation}
where ${\cal N}_Y$, the imaginary part of the forward scattering amplitude, is defined to be ${\cal N}_Y= 1 - {1\over N_c}\langle{\rm Tr}(V_x V_y^\dagger)\rangle_Y$.
Note that the size of the dipole, ${\vec r}_\perp = {\vec x}_\perp - {\vec y}_\perp$ and ${\vec b} = ({\vec x}_\perp + {\vec y}_\perp)/2$. The JIMWLK equation for the two point Wilson correlator is identical in this
large A, large $N_c$ mean field limit to an equation derived independently by Balitsky and Kovchegov-the BK equation~\cite{BK}, which has the operator form
\begin{equation}
{{\partial {\cal N}_Y}\over \partial Y} = {\bar \alpha_S}\, {\cal K}_{\rm BFKL} \otimes \left\{ {\cal N}_Y - {\cal N}_Y^2\right\} \, .
\label{eq:8}
\end{equation}
Here ${\cal K}_{\rm BFKL}$ is the well known BFKL kernel. When ${\cal N} << 1$, the quadratic term is small and one has BFKL growth of the number of dipoles; when ${\cal N}$ is close to unity,
the growth saturates. The approach to unity can be computed analytically~\cite{LevinTuchin}. The BK equation is the simplest equation including both the Bremsstrahlung responsible for the rapid growth of amplitudes at small x as well as the repulsive many body effects that lead to a saturation of this growth.

In this framework, a saturation condition, say ${\cal N} =1/2$, determines the saturation
scale. One obtains  $Q_s^2 = Q_0^2 \exp( \lambda Y)$, where $\lambda = c \alpha_S$ with $c\approx 4.8$. As we shall discuss further in the
next subsection, the (arbitrary) choice of this saturation condition affects the overall normalization of this scale but does not affect the power $\lambda$. In fixed coupling, the power $\lambda$ is large and there are large pre-asymptotic corrections to this relation-which die off only slowly as a function of $Y$. BFKL running coupling effects change the behavior of the saturation scale completely-one goes smoothly at large $Y$ to $Q_s^2 = Q_0^2 \exp(\sqrt{2b_0 c(Y + Y_0}))$ where $b_0$ is the coefficient of the one-loop QCD
$\beta$-function. An impressive computation of $Q_s$ is the work of Triantafyllopoulos, who obtained $Q_s$ by solving
NLO-resummed BFKL in the presence of an absorptive boundary (which approximates the CGC)~\cite{Dionysis2}. The pre-asymptotic effects are much smaller in this case and the coefficient $\lambda\approx 0.25$ is very close to the value extracted from saturation model fits to the HERA data. There is currently an intense, on-going effort to directly compute the NLO corrections to the leading order kernels of the BK equation~\cite{ABKW}.
No analytical solution of the leading order BK equation exists in the entire kinematic region but there have been several numerical studies at both fixed and running coupling~\cite{GSM,Armesto,Albacete}. These studies suggest that the solutions have a soliton like structure and that the saturation scale has the
behavior discussed here. 

The soliton like structure of the numerical solutions is not accidental, as was discovered by Munier and Peschanski~\cite{MunierPeschanski}. They noticed that the BK-equation, in a diffusion approximation, bore a formal analogy to the FKPP equation describing the propagation of unstable non-linear wavefronts in statistical mechanics~\cite{FKPP}. In addition, the full BK-equation lies in the universality class of the FKPP equation. This  enables one to extract the universal properties such as the leading pre-asymptotic terms in the expression for the saturation scale. It was realized~\cite{IMMunier} that a stochastic generalization of the FKPP equation-the sFKPP equation-could provide insights into impact parameter dependent fluctuations~\cite{MuellerShoshi} in high energy QCD beyond the BK-equation. We shall not discuss further efforts in that direction here except to note that there are many open ends (and opportunities) here which are still not satisfactorily resolved and require new ideas.

\subsection{Color Dipole CGC models in DIS and hadronic scattering}
\begin{figure}[htbp]
%\begin{minipage}[t]{.2\linewidth}
\begin{center}
\includegraphics[width=0.32\linewidth]{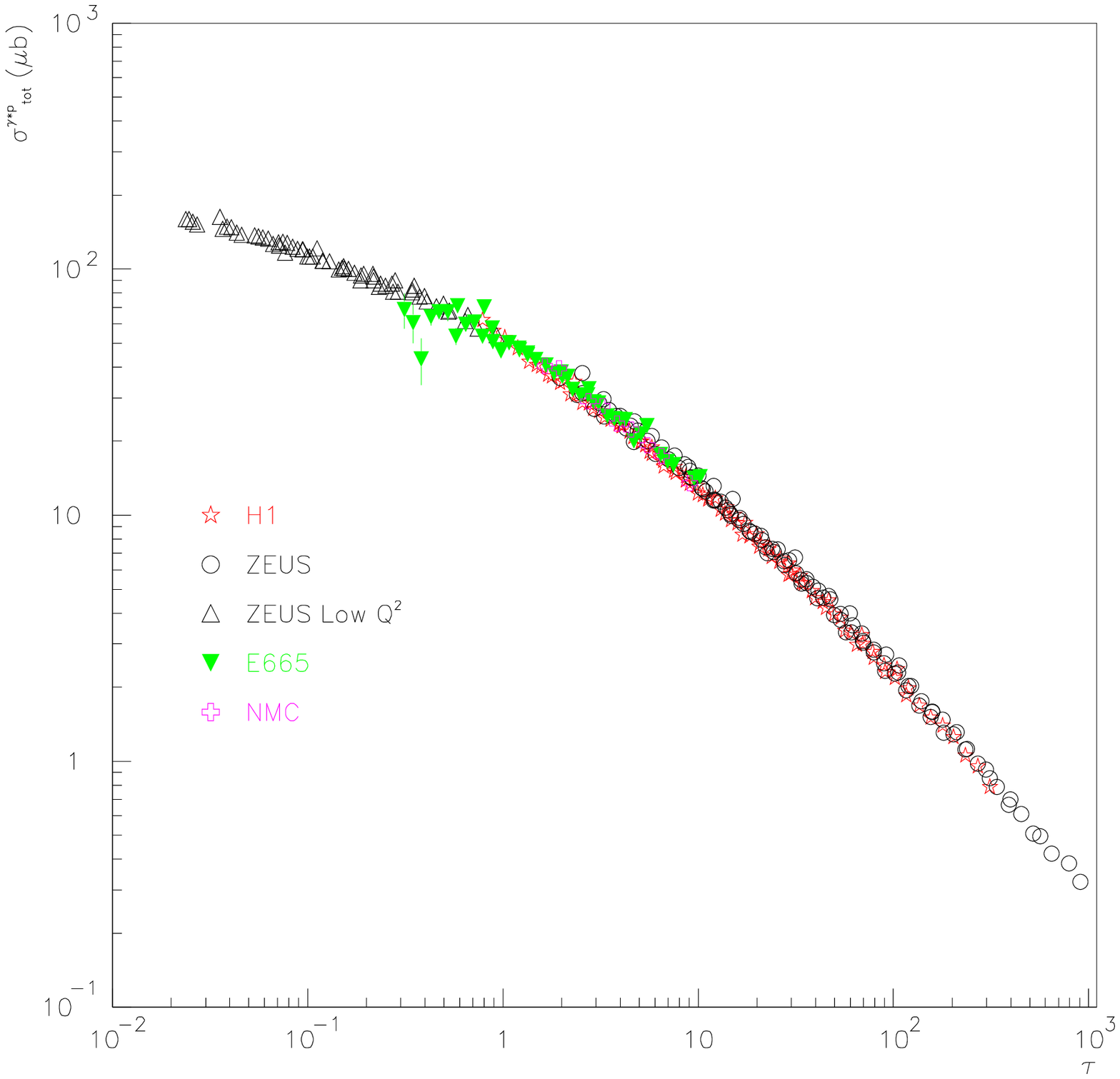}
%\end{center}
%\end{minipage}
%\begin{minipage}[t]{.2\linewidth}
%\begin{center}
\includegraphics[width=0.32\linewidth]{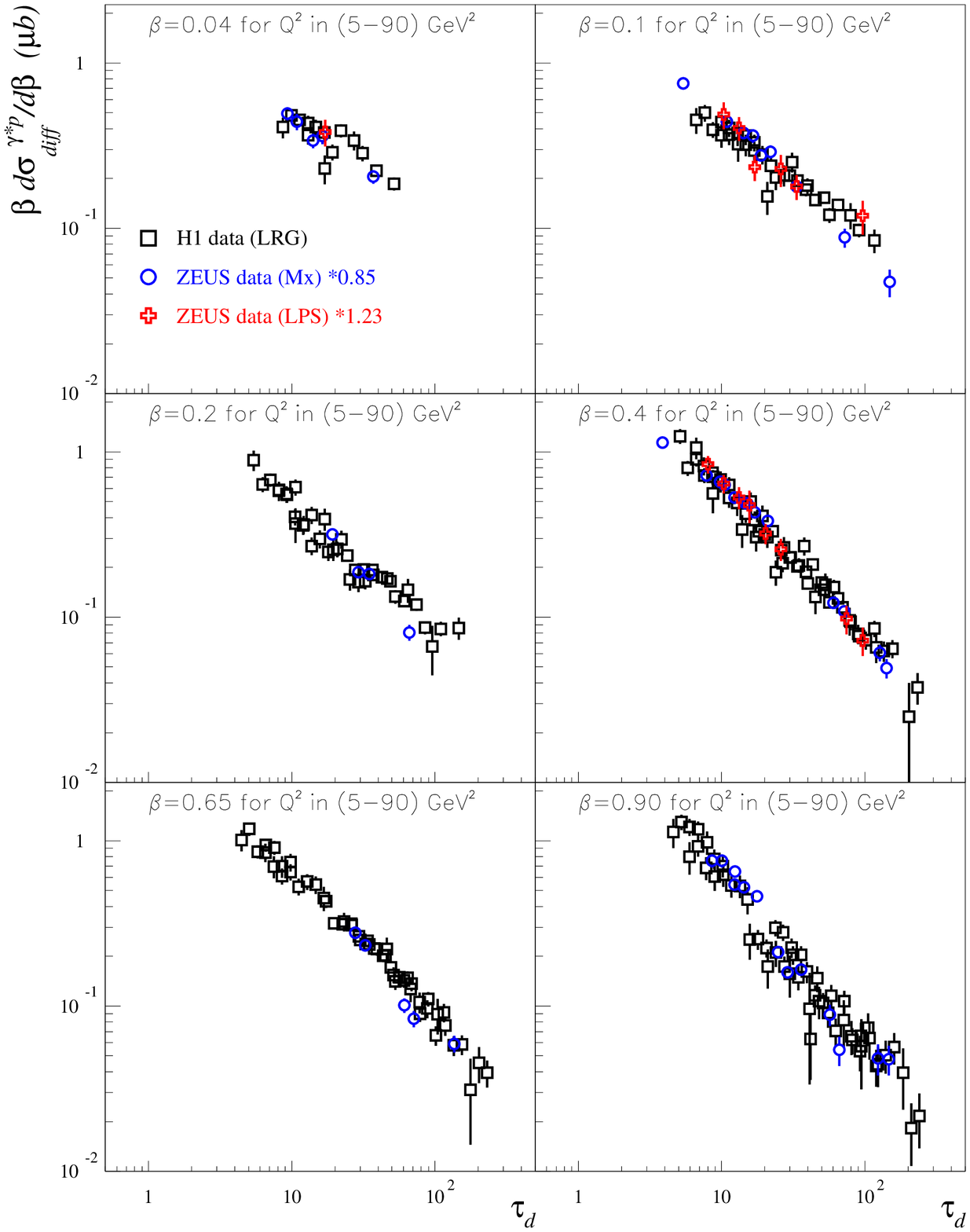}
%\end{center}
%\end{minipage}
%\begin{minipage}[t]{.2\linewidth}
%\begin{center}
\includegraphics[width=0.32\linewidth]{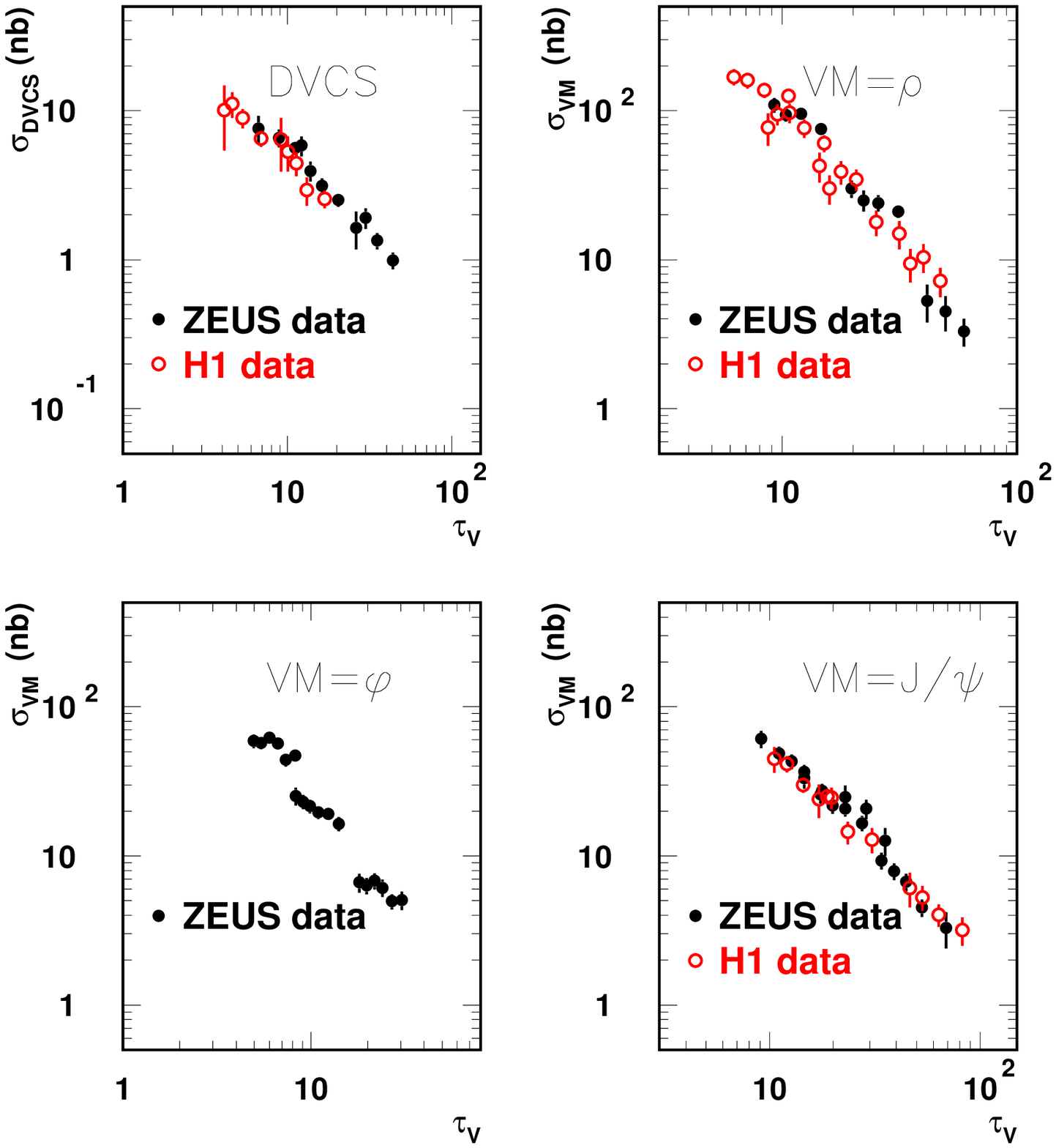}
\end{center}
%\end{minipage}
\caption{Geometrical scaling fully inclusive, diffractive and exclusive vector meson cross-sections. From ~\cite{MarquetSchoeffel}.}
\label{fig:scaling}
\end{figure}

In the previous sub-section, we discussed the formalism of the CGC and the JIMWLK/BK RG equations. We will here discuss phenomenological applications
of these ideas to DIS and (more briefly) to hadronic collisions. The next lecture will discuss more at length the application of these ideas to heavy ion collisions.
%%%%%%%%%%%%%%%%%%%%%%%%%%%%%%%%%%%%%%%%%%%
\begin{figure}
\centering
\includegraphics[scale=0.5]{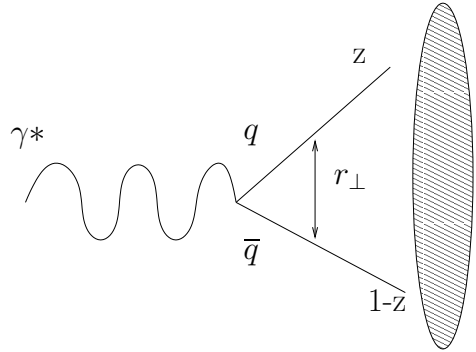}
\caption{Interaction of the color dipole with the nuclei}
\label{fig:dipole}       % Give a unique label
\end{figure}
%%%%%%%%%%%%%%%%%%%%%%%%%%%%%%%%%%%%%%%%%%
A strong hint that semi-hard scales may play a role in small x dynamics came from ``geometrical scaling" of the HERA data~\cite{stasto}. The inclusive virtual photon+proton cross-section for $x\le 0.01$ and
all available $Q^2$ scales~\footnote{The E665 data are a notable exception.} as a function of $\tau\equiv Q^2/Q_s^2$, where $Q_s^2(x) = \exp(\lambda Y)$ GeV$^2$. Here $Y=\ln(x_0/x)$ is the rapidity;  $x_0= 3\cdot 10^{-4}$ and $\lambda= 0.288$ are parameters fit to the data~\cite{stasto,MarquetSchoeffel}. Geometrical scaling of the e+p data is shown in fig.~\ref{fig:scaling}, which demonstrates that the inclusive diffractive, vector meson and DVCS cross-sections at HERA, with a slight modification\footnote{$\tau_{D,VM} = (Q^2 + M^2)/Q_s^2$, where $M$ denotes the mass of the diffractive/vector meson final state.}in the definition of $\tau$, also appear to show geometrical scaling~\cite{MarquetSchoeffel}.   A recent ``quality factor" statistical analysis~\cite{Gelis-etal} indicates that this scaling is robust; it is however unable to distinguish between the above fixed coupling energy dependence of $Q_s$ and the running coupling $Q_s(x)\propto \exp(\sqrt{Y})$ dependence of the saturation scale.
%%%%%%%%%%%%%%%%%%%%%%%%%%%%%%%%%%%%%%%%%%
%%%%%%%%%%%%%%%%%%%%%%%%%%%%%%%%%%%%%%%%%
Geometrical scaling is only asymptotic in both fixed and running coupling evolution equations.  However, recent analyses~\cite{AlbaceteKovchegov,Cyrille} suggests that the onset of the scaling asymptotics may be precocious. Geometrical scaling alone is not sufficient evidence of saturation effects and it is important to look at the data in greater detail in saturation/CGC models.

All saturation models~\cite{Mueller1989:st} express the inclusive virtual photon+proton cross-section as
%%%%%%%%%%%%%%%%%%%%%%%%%%%%%
\begin{equation}\label{eq:sigmatot}
\sigma^{\gamma^*p}_{L,T}%(x,Q^2) 
 = \int   d^2 r_\perp  \int_0^1  d z \left| \Psi^{\gamma^*}_{L,T} (r_\perp,Q,z) \, .
  \right|^2 \, \sigma_{q\bar q P}(r_\perp,x_{\rm Bj}, b_\perp)
\end{equation}
%%%%%%%%%%%%%%%%%%%%%%%%%%%%%%%
Here $\left| \Psi_{L,T}^{\gamma^*}(r_\perp,z,Q) \right|^2$
represents the probability for a  virtual photon to produce a quark--anti-quark pair of size $r = |r_\perp|$ and $\dsigmap(r_\perp,x_{\rm Bj},b_\perp)$ denotes the \emph{dipole cross section} for this pair to scatter off the target at an impact parameter $b_\perp$. The former is well known from QED, while the latter represents the dynamics of QCD scattering at small x--see eq.~\ref{eq:sigmaqqP}. A simple saturation model
(known as the GBW model~\cite{Golec-Biernat:1998js})
of the dipole cross section, parametrized as $\sigma_{q\bar q P} = 2 ( 1 - e^{ - r^2 Q_{s,p}(x)}/4)$ where
$Q_{s,p} (x) = (x_0/x)^\lambda\, \mbox{GeV}^2$, gives a good qualitative fit to the HERA inclusive cross section data for
$x_0 = 3\cdot 10^{-4}$ and $\lambda = 0.288$. Though this model captures the qualitative features of saturation, it does not contain the bremsstrahlung limit of perturbative QCD (pQCD) that applies to small dipoles of size 
$r \ll 1/Q_{s,p}(x)$. 

 In the classical effective theory of the
 CGC, one can derive, to leading logarithmic accuracy, the dipole cross section~\cite{Venugopalan:1999wu}
 containing the right small $r$ limit. This dipole cross section can be
 represented  as~\cite{Kowalski:2003hm}
%%%%%%%%%%%%%%%%%%%%%%%%%%%%%%
 \begin{equation}
 \sigma_{q\bar q P}(r_\perp,x_{\rm Bj}, b_\perp)
  = 2\,\left[ 1 - \exp\left(- r^2  F({\rm x_{Bj}},r_\perp) T_p(b_\perp)\right)
 \right],
 \label{eq:BEKW}
 \end{equation}
%%%%%%%%%%%%%%%%%%%%%%%%%%%%%%
 where $T_p(b_\perp)$ is the impact parameter profile function in the proton, normalized as
 $\int d^2 {\vec b} \,T_p(b_\perp) = 1$ and $F$ is proportional to the gluon distribution~\cite{Bartels:2002cj}: 
%%%%%%%%%%%%%%%%%%%%%%%%%%%%
 \begin{equation}
 F(x_{\rm Bj},r_\perp^2) = \pi^2 \alpha_S \left(\mu_0^2 + 4/r_\perp^2 \right)
 x_{\rm Bj} g\left(x_{\rm Bj},\mu_0^2 + 4/r_\perp^2 \right)/(2 N_c)\,,
 \label{eq:BEKW_F}
 \end{equation}
%%%%%%%%%%%%%%%%%%%%%%%%%%%%%
 evolved from the initial scale $\mu_0$ by the DGLAP equations. The dipole cross section in eqn. \ref{eq:BEKW} was
 implemented in the impact parameter saturation model (IPsat)~\cite{Kowalski:2003hm} where the parameters are fit to
 reproduce the HERA data on the inclusive structure function $F_2$. Here $Q_{s,p}$ 
is defined\footnote{This choice of  is equivalent to the
 saturation scale in the GBW model for the case of a Gaussian dipole cross section.} as the solution of $ \sigma_{q\bar q P}(r_\perp,x_{\rm Bj}, b_\perp)= 1/Q_{s,p}(x_{\rm Bj},b_\perp) = 2(1-e^{-1/4})$.

The IPsat dipole cross section in eqn. \ref{eq:BEKW} is valid when leading logarithms in
x in pQCD are not dominant over leading logs in $Q^2$. At very small x, where
logs in x dominate, quantum evolution in the CGC describes both the BFKL limit of linear small x evolution as well as nonlinear JIMWLK/BK evolution at high parton densities~\cite{JIMWLK,BK}.
These asymptotics are combined with a more realistic $b$-dependence
in the b-CGC model~\cite{Iancu:2003ge,Kowalski:2006hc}. Both the IPsat model and 
the b-CGC model provide excellent fits to HERA data for $x \leq 0.01$~\cite{Kowalski:2006hc,Forshaw:2006np}.
%%%%%%%%%%%%%%%%%%%%%%%%%%%%%
\begin{figure}[htbp]
\begin{center}
\resizebox*{6cm}{!}{\includegraphics{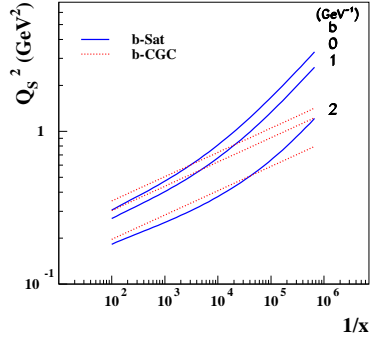}}
\end{center}
\caption{\label{fig:Qs} The saturation scale vs $1/x$ in the IPsat and b-CGC models~\cite{Kowalski:2006hc}.}
\end{figure}
%%%%%%%%%%%%%%%%%%%%%%%%%%%%%%
The saturation scale extracted from the fit in the IPsat model is shown in Fig.~\ref{fig:Qs}. The important point to note is that 
the energy dependence of the extracted $Q_{s,p}$ is significantly stronger than those predicted in non-perturbative models~\cite{DonnachieLandshoff}.

\begin{figure}[htbp]
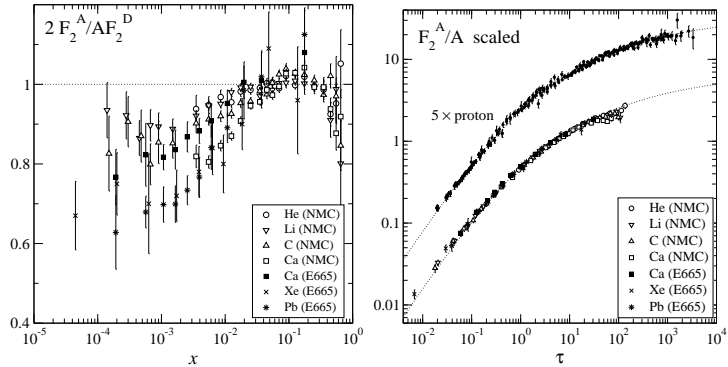

%\begin{minipage}[t]{.3\linewidth}
\begin{center}
\includegraphics[width=0.4\linewidth]{ratio_all_notscaled_bw.eps}
%\end{center}
%\end{minipage}
%\begin{minipage}[t]{.3\linewidth}
%\begin{center}
\includegraphics[width=0.4\linewidth]{F2_N+P_scaled_bw.eps}
\end{center}
%\end{minipage}
\caption{Left: Shadowing of $F_2$ from the NMC and E665 fixed target experiments. Right: The data scaled as a
function of $\tau\equiv Q^2/Q_{s,A}^2$~\cite{Freund:2002ux}.}
\label{fig:shad}
\end{figure}

\begin{figure}[htbp]
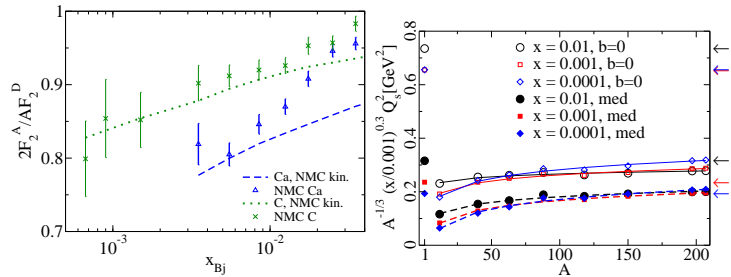

%\begin{minipage}[t]{.3\linewidth}
\begin{center}
\includegraphics[width=.4\linewidth]{glaubercompilationnmc.eps}
%\end{center}
%\end{minipage}
%\begin{minipage}[t]{.3\linewidth}
%\begin{center}
\includegraphics[width=.4\linewidth]{Qscombination.eps}
\end{center}
%\end{minipage}
\caption{Left: Comparison of the IPsat model (with no adjustable parameters) to the NMC data. Right:  The
$A$ and x dependence of the saturation scale in the IPsat model~\cite{HTR}.}
\label{fig:shad2}
\end{figure}

The strong field dynamics of small x partons is universal and  should be manifest in
large nuclei at lower energies than in the proton. In Fig.~\ref{fig:shad} (left), we show
the well known shadowing of $F_2^A$ in the fixed target e+A E665 and NMC experiments. Expressed in terms of
$\tau\equiv Q^2/Q_s^2$ (Fig.~\ref{fig:shad} (right)), the data show geometrical scaling~\cite{Freund:2002ux}. 
%%%%%%%%%%%%%%%%%%%%%%%%%%%
%%%%%%%%%%%%%%%%%%%%%%%%%%%%%%
A careful study of nuclear DIS in the IPsat CGC framework was performed in Ref.~\cite{Kowalski:2003hm,HTR}. The
average differential dipole cross section is well approximated by $\langle{d\sigma_A\over d^2 b_\perp}\rangle_N \approx 2\left[1-\left(1-\frac{T_A(b_\perp)}{2}\sigma_{p} \right)^A\right]$, where $T_A(b_\perp)$ is the well known Woods Saxon distribution. The average is defined as
$\langle{\mathcal{O}}\rangle_N \equiv \int \prod_{i=1}^A d^2 b_{\perp,i} T_A(b_{\perp,i}) \mathcal{O}(\left\{b_{\perp,i}\right\})$. Here $\sigma_p$ 
is determined from the IPsat fits to the e+p data; no additional parameters are introduced for $eA$ collisions. In Fig.~\ref{fig:shad2} (left), the model is compared to NMC data on Carbon and Calcium nuclei-the agreement is quite good. In Fig.~\ref{fig:shad2} (right), we show the extracted
saturation scale in nuclei for both central and median impact parameters. To a good approximation,  the saturation scale in nuclei scales as $Q_{s,A}^2(x,b_{\rm med.}) \approx Q_{s,p}^2(x,b_{\rm med.})\cdot (A/x)^{1/3}$.  The factor of $200^{1/3} \approx 6$ gives a huge ``oomph'' in the parton density of a nucleus relative to that of a proton at the same x. \emph{Indeed, one would require a center of mass energy 
$\sim 14$ times larger in the proton case}. At extremely high energies, this statement must be qualified to account for the
effects of QCD evolution~\cite{Mueller:2003bz}.
%%%%%%%%%%%%%%%%%%%%%%%%%%%%%%%%%%%%%

We now turn to a discussion of CGC effects in hadronic collisions. A systematic
treatment of the scattering of two strong color sources (such as two high energy nuclei) is discussed in the next lecture. To leading order, the problem reduces to the solution of the classical Yang-Mills (CYM) equations averaged over color sources for each nucleus~\cite{KovnerMW,KV}; the variance of this distribution of sources is proportional to $Q_{s,A}^2$. Besides the nuclear radius, it is the only scale in the problem, and
the  $Q_{s,A}^2 \sim Q_{s,}^2 \cdot (A/x)^{0.3}$ expression for the saturation scale was used in CGC models of nuclear
collisions to successfully predict the multiplicity~\cite{KV} and the centrality dependence of the multiplicity~\cite{KLN} dependence in gold+gold collisions at RHIC. The universality of the saturation scale also has a bearing on the hydrodynamics of the Quark Gluon Plasma (QGP); the universal form leads to a lower eccentricity~\cite{LappiV} (and therefore lower viscosity) than a non-universal form that generates a larger eccentricity~\cite{Hirano-etal} (leaving room for a larger viscosity) of the QGP\footnote{We shall present yet another take on this issue in lecture II.}

\begin{figure}[htbp]
\begin{center}
\resizebox*{6cm}{!}{\includegraphics{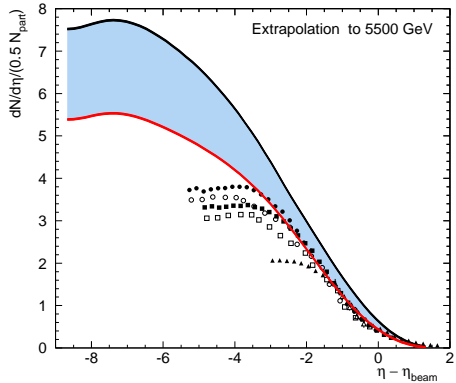}}
\end{center}
\caption{\label{fig:LF}Prediction for limiting fragmentation
and deviations away from it at LHC energies. The bands denote the range in the predictions for GBW and MV models. From
~\cite{FAR}. }
\end{figure}

For asymmetric (off-central rapidity) nuclear collisions, or proton/deuteron + heavy nucleus collisions, $k_\perp$-factorization
can be derived systematically for gluon production, at leading order, in the CGC framework~\cite{KovchegovMueller}. 
Limiting fragmentation~\cite{Jalilian} and deviations thereof, are described by solutions of the BK-equation. Predictions for the multiplicity distribution in A+A collisions at the LHC~\cite{FAR} for both Golec--Biernat--Wusthoff (GBW)  and classical CGC (MV) dipole initial conditions~\footnote{The MV initial condition has the same form as the IPsat dipole cross-section discussed earlier.} give a charged particle multiplicity of 1000-1400 in central lead+lead collisions at the LHC\footnote{See Ref.~\cite{Armesto-QM08} for other model predictions.}. The results are shown in Fig.~\ref{fig:LF}.
%%%%%%%%%%%%%%%%%%%%%%%%%%%%%%
%%%%%%%%%%%%%%%%%%%%%%%%%%%%%%%%

\begin{figure}[htbp]
%\begin{minipage}[t]{.3\linewidth}
\begin{center}
%\includegraphics[width=.25\linewidth]{star_pi0.eps}
%\end{center}
%\end{minipage}
%\begin{minipage}[t]{.3\linewidth}
%\begin{center}
\includegraphics[width=.4\linewidth,angle=270]{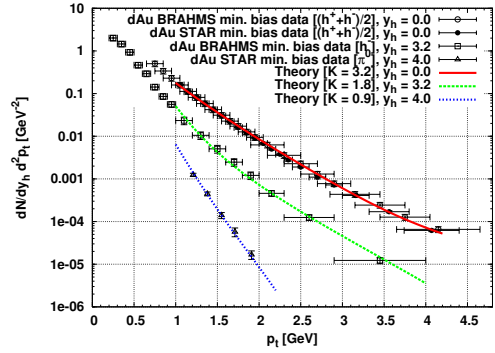}
\end{center}
%\end{minipage}
\caption{The inclusive $k_\perp$ distributions in deuteron+gold collisions compared to theory curves for different
rapidities. From ~\cite{DumitruHJ}.}
\label{fig:RdA}
\end{figure}

In deuteron + gold collisions at RHIC, data on the inclusive hadron spectra\footnote{The same analysis also gives good agreement for the forward
p+p spectrum at RHIC~\cite{BoerDH}.} can be directly compared to model predictions~\cite{DumitruHJ}. The result is shown in Fig.~\ref{fig:RdA}. For a comprehensive review of applications of CGC picture to RHIC phenomenology, we refer the reader to Ref.~\cite{YJ2}.
%%%%%%%%%%%%%%%%%%%%%%%%%%%%%
%%%%%%%%%%%%%%%%%%%%%%%%%%%%%
There are a couple of caveats to this picture. Firstly, $k_\perp$ factorization is very fragile. It does not hold for quark production even at leading order in the parton density~\cite{BlaizotGV2}, albeit it may be a good approximation for large masses and transverse momenta~\cite{GelisV1}. For gluon production, it does not hold beyond leading order in the parton density~\cite{Balitsky2,KV}. Secondly, a combined comprehensive analysis of HERA
and RHIC data is still lacking though there have been first attempts in this direction~\cite{KugeratskiGN1}.
%%%%%%%%%%%%%%%%%%%%%%%%%%%%%%%%%%%%%%%%
\subsection{The future of small x physics at hadron colliders and DIS}

\begin{figure}[htbp]
%\begin{minipage}[t]{.3\linewidth}
\begin{center}
\includegraphics[width=0.4\linewidth]{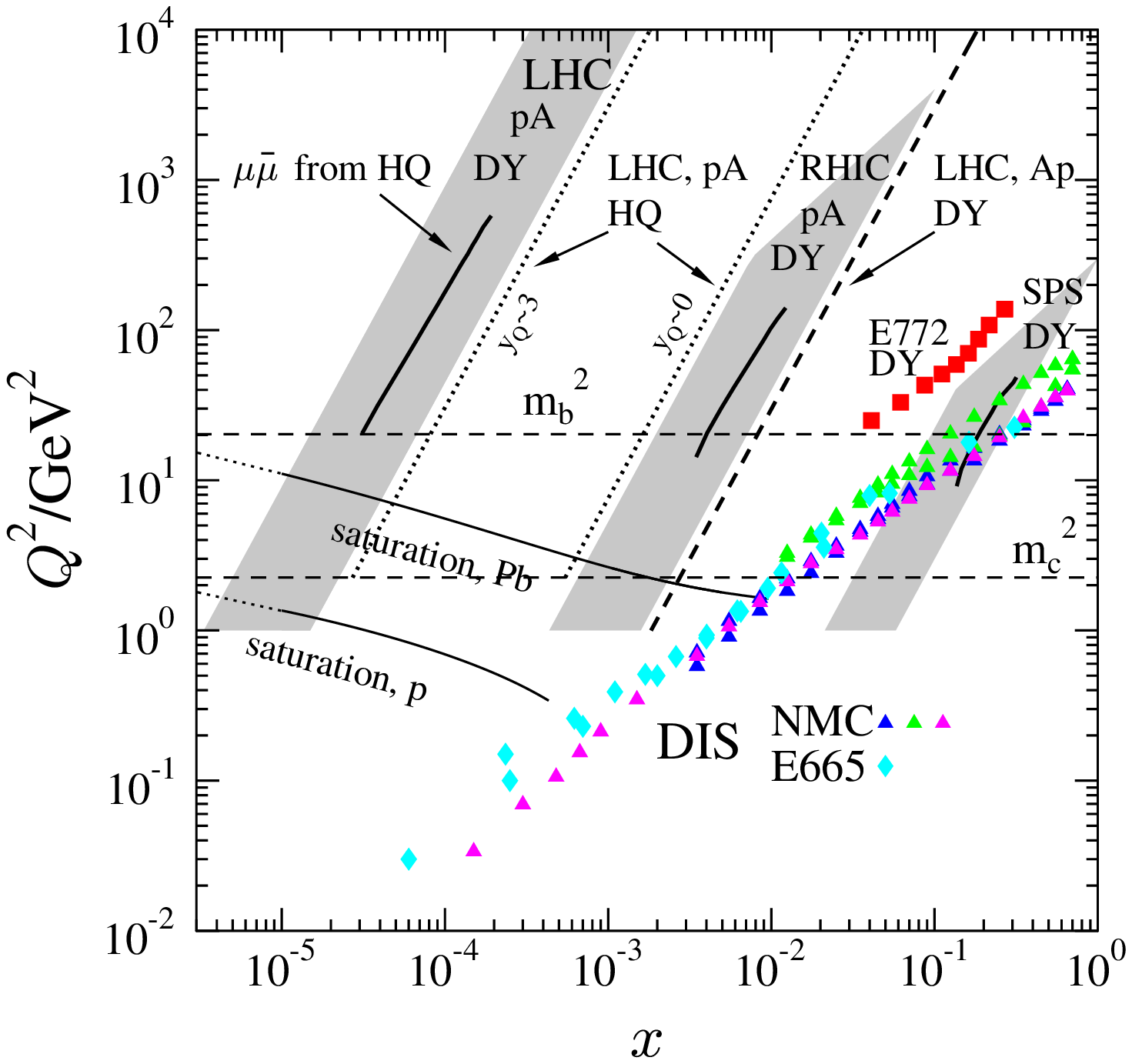}
%\end{center}
%\end{minipage}
%\begin{minipage}[t]{.3\linewidth}
%\begin{center}
\includegraphics[width=0.4\linewidth]{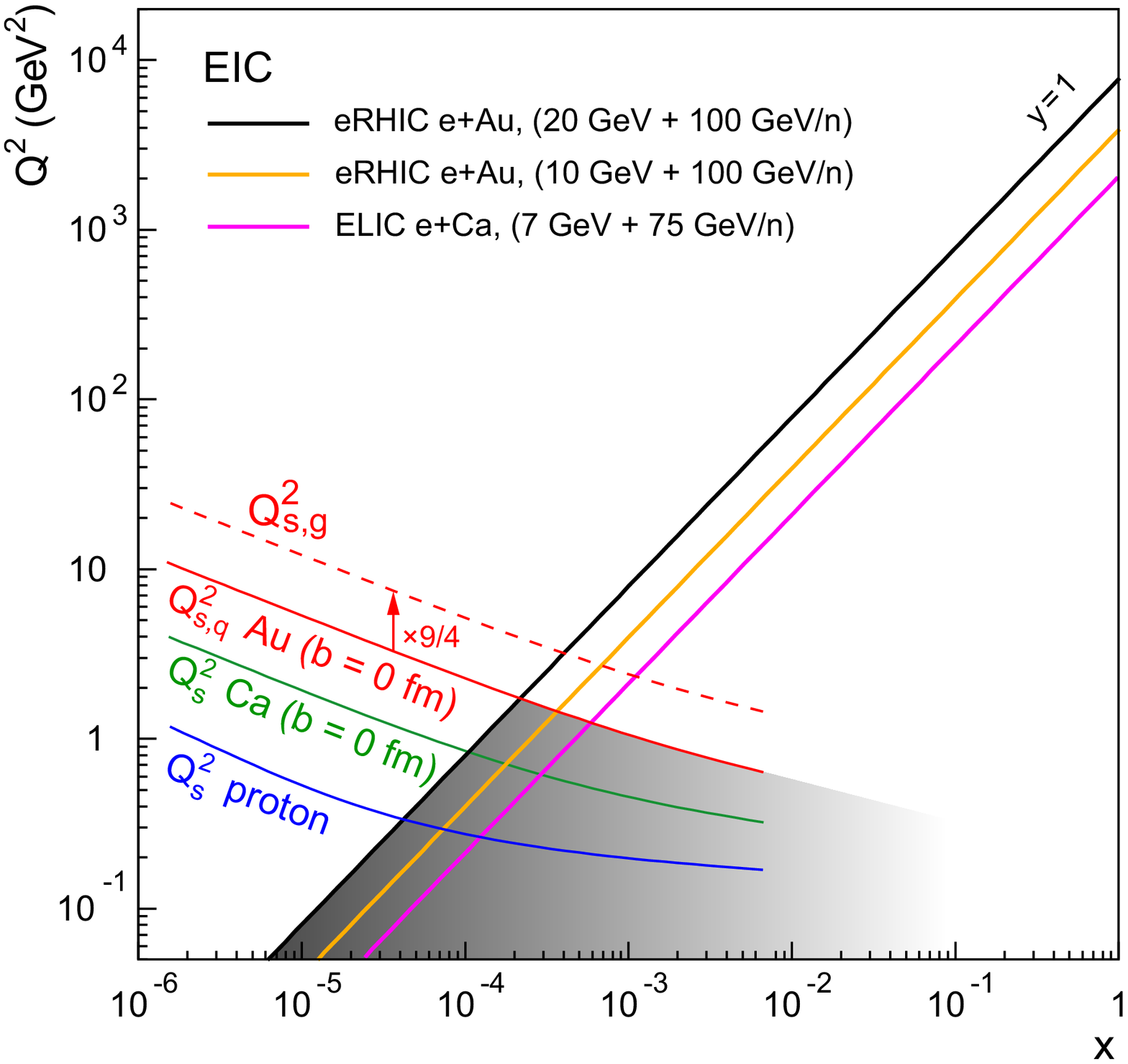}
\end{center}
%\end{minipage}
\caption{Left: Kinematic x-$Q^2$ reach of different final states at the LHC compared to other experiments with
nuclei . From ~\cite{D'Enterria}. Right: The saturation
scale in the proton, calcium and gold in the kinematic acceptance of the EIC.}
\label{fig:LHC-EIC}
\end{figure}

The LHC is the ultimate small x machine in terms of reach in x for large $Q^2$. A plot from
Ref.~\cite{D'Enterria} illustrating this reach is shown in Fig.~\ref{fig:LHC-EIC} (left). For a recent review of the small x opportunities at the LHC, see Ref.~\cite{WeissFS}. The LHC will provide further, more extensive tests of the hints for the CGC seen at RHIC.
The universality of parton distributions is often taken for granted but factorization theorems proving this universality have been proven only for a limited number of inclusive final states. However, as we have discussed, small x is the domain of rich multi-parton correlations. These are more sensitive to more exclusive final states for which universality is not proven~\cite{CollinsQiu}. Therefore, while the LHC will have unprecedented reach in x, precision studies of high energy QCD and clean theoretical interpretations of these motivate future DIS projects. Two such projects
are the EIC project in the United States~\cite{Surrow} and the LHeC project in Europe~\cite{Newman}.
%%%%%%%%%%%%%%%%%%%%%%%%%%%%%%
%%%%%%%%%%%%%%%%%%%%%%%%%%%%%%%%%%%%%%%%%
As we discussed previously, strong color fields may be more easily accessible in DIS off nuclei relative to the proton due to the ``oomph" factor. In Fig.~\ref{fig:LHC-EIC} (right), we show the saturation scale $Q_{s,A}(x)$ overlaid on the x-$Q^2$ kinematic domain spanned by the EIC. As suggested by the figure, the EIC (and clearly the higher energy LHeC) will cleanly probe the cross-over regime from
linear to non-linear dynamics in QCD. For further discussion of the physics of an Electron Ion collider, see Ref.~\cite{ARRW,Ullrich}.
%%%%%%%%%%%%%%%%%%%%%%%%%%%%%%%%%%%%%%%%%%

\section*{Heavy Ion Collisions}
\begin{figure}
\centering
\includegraphics[scale=0.45,angle=270]{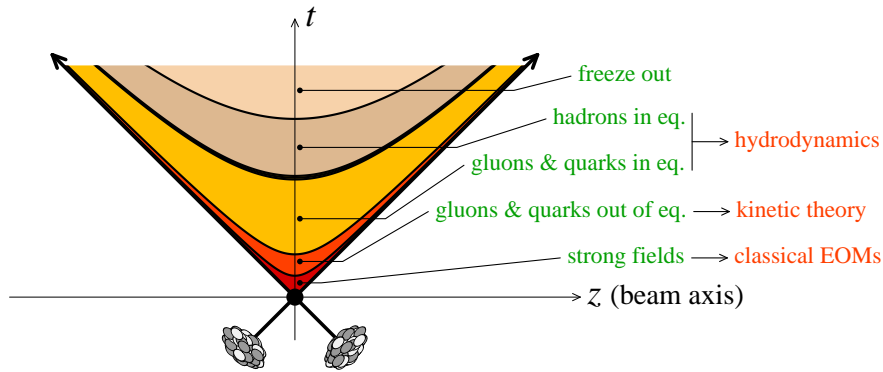}
\caption{The various stages of a heavy ion collision.}
\label{fig:spacetime}       % Give a unique label
\end{figure}

In the first lecture, we discussed the physics of high parton densities in QCD. We motivated a description of the high energy structure of hadrons and nuclei
as a Color Glass Condensate. We discussed the phenomenological application of CGC based models to describe data on DIS and  hadronic collisions.
In this lecture, we will focus our attention on heavy ion collisions and try and understand {\it ab initio} the properties of the QCD matter that is formed when 
two high energy nuclear wavefunctions--sheets of colored glass--collide. A schematic space--time picture of the evolution of a heavy ion collision is
shown in fig.~\ref{fig:spacetime}. 

We shall not attempt to describe all features of a heavy ion collision but merely focus on the very early initial dynamics for its intrinsic interest but 
also because it is important to understand to determine whether the matter thermalizes and can be subsequently described by hydrodynamics. The QCD matter that is formed at very early times is a coherent classical field, which expands, decays into nearly on shell partons and may eventually thermalize to form a Quark Gluon Plasma (QGP).  Because it is formed by melting the frozen CGC degrees of freedom, and because it is the non-equilibrium matter preceding the QGP, this matter is called the Glasma~\cite{Glasma}.

%%%%%%%%%%%%%%%%%%%%%%%%%%%%
%%%%%%%%%%%%%%%%%%%%%%%%%%%%%%%%%%%
\section{From CGC to Glasma}
\label{sect:three}
%%%%%%%%%%
We will first discuss the classical picture of the Glasma that emerges from solutions of classical Yang--Mills equations. We will then
discuss the role of quantum corrections in the Glasma.

\subsection{The Classical Solution}
%%%%%%%%%%%%%%%%%%%%5
Let us begin by setting up the kinematics involved for the problem of two ultra-relativistic nuclei approaching each other. They are highly Lorentz contracted and can be considered to be sitting at $x^{\pm}=0$ for $t < 0$ in the light cone coordinates. They collide at $x=t=0$. At $t \geq 0$ it is convenient to introduce the proper time $\tau=\sqrt{t^2-z^2}$ which is invariant under Lorentz boosts and the space time rapidity
%%%%%%%%%%%%%%%%%%%%%%%%%%%%%%%
\begin{equation}
 \eta = \frac{1}{2} \ln \left( \frac{t+z}{t-z} \right)
\end{equation}
%%%%%%%%%%%%%%%%%%%%%%%%%%%%%%%%%
For free streaming particles $z=vt=p_zt/E$ and the space-time rapidity equals the momentum space rapidity y, namely, $\eta=y$.
%%%%%%%%%%%%%%%%%%%%%%%%%%%%%%%%
\begin{figure}
\centering
\includegraphics[scale=0.3,angle=270]{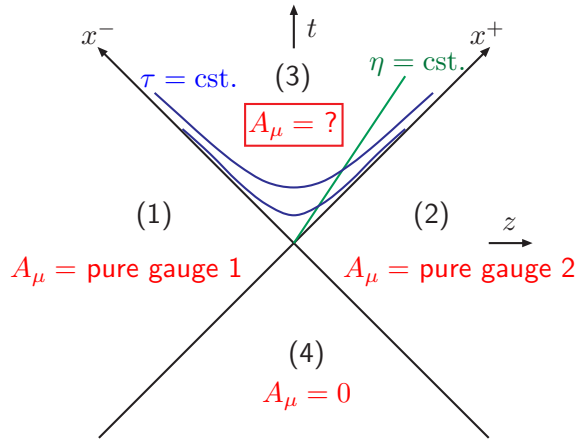}
\caption{The space-time distribution of gauge fields before and after the collision.}
\label{fig:potentials}       % Give a unique label
\end{figure}
%%%%%%%%%%%%%%%%%%%%%%%%%%

Let us make some some approximations to get to the solutions without sacrificing any essential physics. The collisions are considered to be at very high energy so that the nuclei are static light cone currents that are delta functions in $x^\mp$ respectively for nuclei whose large momentum component is
given by $P^\pm$. A consequence is that the solutions of the classical Yang-Mills equations, for this source distribution, are boost invariant. This leads to a considerable simplification since the equations now only depend on the two transverse directions and the proper time.
%%%%%%%%%%%%%%%%%%%%%%%%%%
\begin{figure}
\centering
\includegraphics[scale=0.3,angle=270]{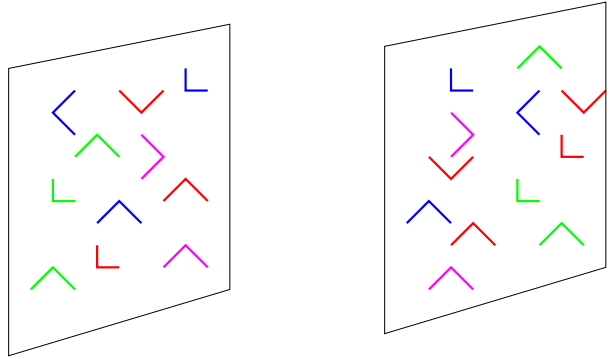}
\caption{The color electric and the magnetic fields lying on the sheets of the charge distribution. They are perpendicular to each other and to the beam direction.}
\label{fig:bfrcoll}       % Give a unique label
\end{figure}
%%%%%%%%%%%%%%%%%%%%%%%%%%%%%%
The initial conditions for the evolution of the classical gauge fields in the forward light cone is illustrated in fig. \ref{fig:potentials}. The fields of the nuclei before the collision are the Li\'{e}nard-Wiechert potentials associated with the respective color charge densities. Both the charge densities and the fields only exist on the sheets and for each source of charge the electric field \vec{E} and the magnetic field \vec{B} are orthogonal to each other and to the beam direction. The situation before the collision is depicted in fig \ref{fig:bfrcoll}. However the vector potentials lie outside the sheets but possess a discontinuity across the sheet according to Gauss law. This is where the small-x gluons are located having a large longitudinal extent and coupling with a host of color sources. As has been argued before they are represented by the classical color fields, have very small lifetime and see the color sources to be static during their lifetime. As noted previously, the infinitesimal nature of the sheets is intended to simplify the problem. A finite size can be taken care of using RG arguments.  Fig. \ref{fig:potentials} shows that the fields vanish in the backward light cone and are two dimensional pure gauge transforms of vacuum in both nuclei before the collision.

\begin{figure}
\centering
\includegraphics[scale=0.3,angle=270]{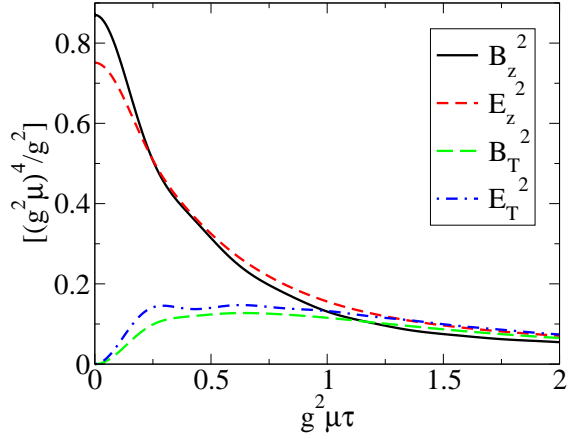}
\caption{The longitudinal and the transverse components of the chromoelectric and the chromomagnetic fields computed numerically~\cite{Glasma}.}
\label{fig:numsolycn}       % Give a unique label
\end{figure}

In the forward light cone, a pure gauge solution of the Yang--Mills (YM) equations of motion cannot be found--the sum of two pure gauges in QCD is not a pure gauge. Solving the YM equations near the light cone \cite{KV,KNV,Lappi}, with the pure gauge initial conditions from the two nuclei before the collision, one finds that the transverse color \vec{E} and \vec{B} fields vanish as $\tau \rightarrow 0$ but the longitudinal fields are non-vanishing. The results of a numerical simulation~\cite{Lappi} are shown in fig.~\ref{fig:numsolycn}.
%%%%%%%%%%%%%%%%%%%%%%%%%%%%%%%%%%%%
%%%%%%%%%%%%%%%%%%%%%%%%%%%%%%%%%%%%%
We note that the non-zero \vec{E} and \vec{B} fields imply that the initial conditions~\cite{KharzeevKV} may have a large density of topological charge $F^{\mu \nu} \tilde{F}_{\mu \nu}$. The LO picture of the Glasma that emerges is one of color flux tubes carrying non-trivial topological charge, localized in the transverse plane and of transverse size $1/Q_s^2$, stretching between the valence color degrees of freedom. As we shall soon see, this picture provides a plausible explanation of the near side ridge in heavy ion collisions.

For the inclusive gluon distribution, the leading order (LO) classical contribution is of order $O(\frac{1}{g^2})$ but all orders in $g\rho_{1,2}$, where $\rho_{1,2}$ are the local color charge densities of the two nuclei. It can be expressed as
%%%%%%%%%%%%%%%%%%%%%%%%%%%%%%%%
\begin{eqnarray}
E_p\frac{dN}{d^3 p}&=&\frac{1}{16\pi^3}
\lim_{x_0,y_0\to+\infty}\int d^3 x d^3 y
 \;e^{ip\cdot(x-y)}
 \;(\partial_x^0-iE_p)(\partial_y^0+iE_p)
 \nonumber\\
 &&\qquad\qquad\times\sum_{\lambda}
 \epsilon_\lambda^\mu(p)\epsilon_\lambda^\nu(p)\;
\big<A_\mu(x)A_\nu(y)\big>\; .
\label{eq:AA}
\end{eqnarray}
%%%%%%%%%%%%%%%%%%%%%%%%%%%%%%%%
The gauge fields on the right hand side are computed numerically for proper times $\tau \geq 0$~\cite{KNV,Lappi} by solving the classical Yang-Mills (CYM) equations in the presence of the light cone current $J^{\mu,a} = \delta^{\mu +} \delta(x^-)\rho_1^a(x_\perp) +\delta^{\mu -}\delta(x^+) \rho_2^a(x_\perp)$. 
The expectation value $\langle \cdots \rangle$ for any inclusive operator ${\cal O}$ denotes\footnote{Here ${\wt {\rho}}$ is the local color charge density 
in Lorentz gauge, which is related by a simple gauge transformation to the corresponding color charge density ${\rho}$ in light cone gauge.We note that 
$W[\rho]$ is gauge invariant. The light cone gauge classical fields are expressed most simply in terms of color charge densities in Lorentz gauge.}
\begin{equation}
\langle {\cal O}\rangle
= \int [D{\wt {\rho}}_1] [D{\wt {\rho}}_2]\;
W_{Y_{\rm beam}-Y}[{\wt{\rho}}_1]\, W_{Y_{\rm beam}+Y}[{\wt {\rho}}_2]\;
{\cal O}\left[{\wt {\rho}}_1,{\wt {\rho}}_2\right] \; .
\label{eq:fact-formula}
\end{equation}
We will justify the validity of this formula shortly. For central rapidity RHIC heavy ion collisions, as discussed in the previous lecture, evolution effects are not important, and $W[{\wt {\rho}}]$ is a Gaussian functional in ${\tilde \rho}$ with the variance $\mu_A^2$. Recall that $Q_s \sim 0.6 \,\mu_A$. So, for these Gaussian distributions, performing the average in eq.~\ref{eq:fact-formula} over the solutions to the Yang--Mills equations in eq.~\ref{eq:AA}, one can compute the number (and energy distributions) of produced gluons in terms of $Q_s$. For the energy density, one obtains $\varepsilon \sim 20$ -- $40$ GeV/fm$^3$ for the values of $Q_s$ we mentioned---obtained by extrapolating from fits to the HERA and fixed target e+A data.

This LO formalism was applied to successfully predict the RHIC multiplicity at $y\sim 0$~\cite{KNV,Lappi} as well as the rapidity and centrality
distribution of the multiplicities~\cite{KLN}. At LO, the initial transverse energy is $E_T \sim Q_S$, which is
about 3 times larger than the final measured $E_T$, while (assuming parton hadron duality) $N_{\rm CGC}\sim N_{\rm had.}$. The two conditions
are consistent if one assumes nearly isentropic flow which reduces $E_T$ due to $PdV$ work while conserving entropy. This assumption has been
implemented directly in ideal hydrodynamic simulations~\cite{Hirano-Nara}.

As discussed previously, CGC based models give values for the initial eccentricity $\epsilon$ that are large than those in Glauber model because the energy and number density locally is  sensitive to the lower of the two saturation scales (or local participant density) in the former and the average of the two in the latter. Naively,  CGC initial conditions would have more flow then and have more room for dissipative effects relative to Glauber. This conclusion is turned on its head in a simple parameterization of incompletely thermalized flow~\cite{BBBO}:
$v_2/\epsilon = {(v_2/\epsilon)_{\rm hydro} \over (1+ K/K_0)}$, where $K = \frac{1}{S_\perp}\sigma \frac{dN}{dy} c_s$ is the Knudsen number, $\sigma$ the
cross-section, $c_s$ the sound speed and $S_\perp$ the transverse overlap area. $K_0$ is a number of order unity. If thermalization were complete,
$K\rightarrow 0$ and one approaches the hydro bound. Computing $\epsilon$ with different initial conditions, and plotting the l.h.s ratio versus $\frac{1}{S_\perp}\frac{dN}{dy}$, one has a two parameter fit to $\sigma$ and $c_s$. The greater CGC eccentricity forces $v_2/\epsilon$ to be lower for more central collisions thereby leading to lower $c_s$;  quicker saturation of $v_2/\epsilon$ forces larger $\sigma$ and therefore lower $\eta$ in the CGC relative to Glauber~\cite{DO}. While is conceivable however that this result may not prove robust against more detailed modeling, it is clear that the results are very sensitive to the initial conditions. 

How much flow is generated in the Glasma before thermalization? The primordial Glasma has
occupation numbers $f\sim \frac{1}{\alpha_S}$ and can be described as a classical field. As the Glasma expands, higher momentum modes increasingly
become particle like and eventually the modes have occupation numbers $f<1$, which may be described by a thermal spectrum.  A first computation of elliptic flow of the Glasma found only about half the observed elliptic flow~\cite{KV-PLB} albeit the computation did not properly treat the interaction between hard and soft modes in the Glasma. Formulating a kinetic theory that describes this evolution is a challenging problem in heavy ion collisions--for a preliminary discussion, see ~\cite{Jeon}.
%%%%%%%%%%%%%%%%%%%%%%%%%%%%%%
\begin{figure}[htb]
\hfill
\includegraphics[width=0.25\textwidth]{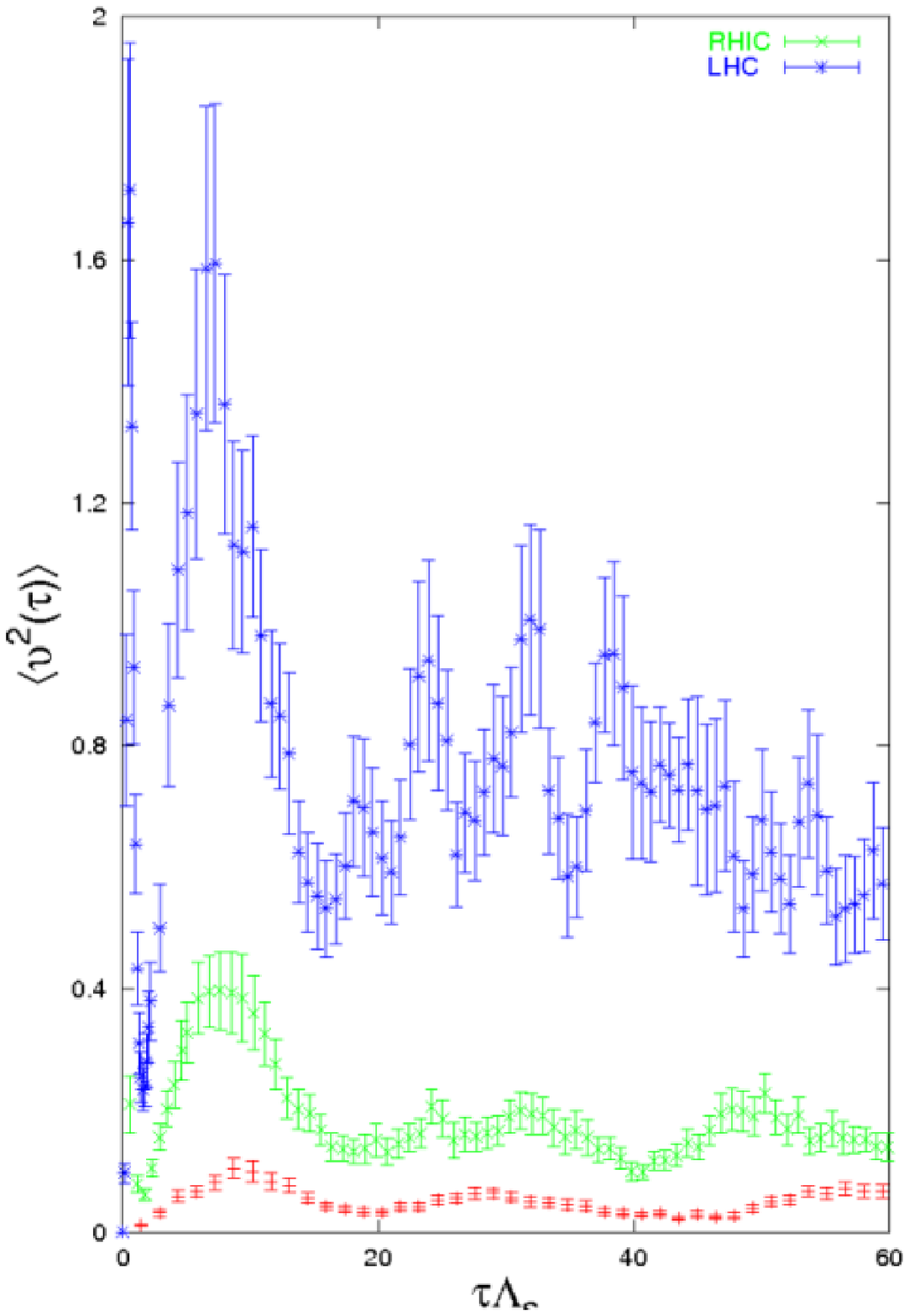}
\hfill
\includegraphics[width=0.5\textwidth]{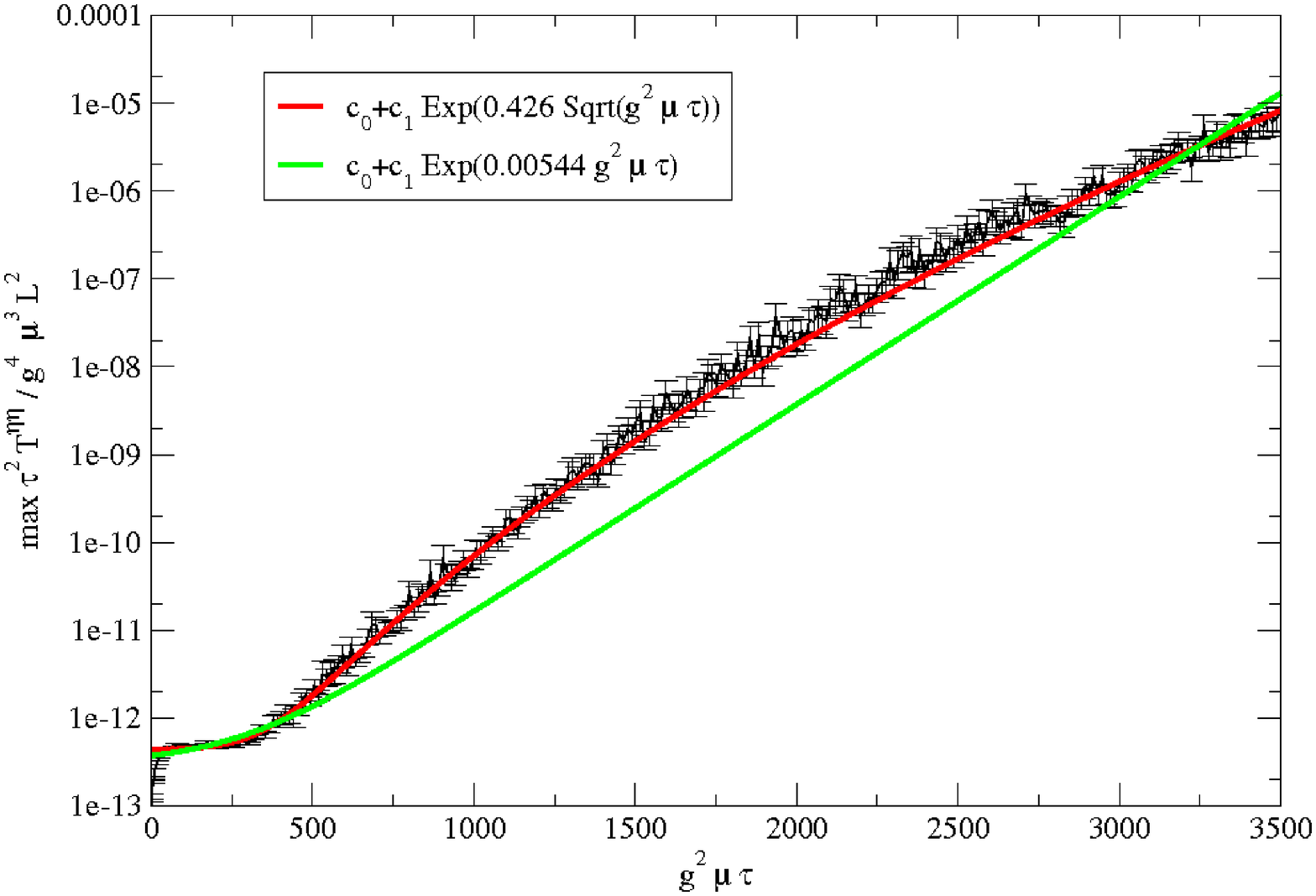}
\hfill
\caption{Left: Chern--Simons mean squared charge as a function of proper time (in units of the saturation scale), generated in the LO boost-invariant 2+1-D Glasma~\cite{KharzeevKV}. Quantum fluctuations allow sphaleron transitions which may give rise to significantly greater values of $\sqrt{\langle \nu^2\rangle}$. Right: The
same quantum fluctuations, albeit suppressed by $\alpha_S$, grow as $\alpha_S\exp(\sqrt{Q_S\tau})$~\cite{PaulR}. These unstable fluctuations,
resummed to all orders, likely give rise to more isotropic initial distributions in the Glasma~\cite{GelisLV1}.}
\label{fig:LO}
\end{figure}

The LO field configurations are unstable and lead to very anisotropic momentum distributions at later times. Such distributions can trigger an instability analogous to the Weibel instability in QED plasmas~\cite{Stan}. It is observed in 3+1-D numerical solutions of CYM equations~\cite{PaulR} that small rapidity dependent quantum fluctuations in the initial conditions generate transverse $E$ and $B$ fields  that grow rapidly as $\exp(\sqrt{Q_S\tau})$. They are the same size as the rapidly diluting longitudinal $E$ and $B$ fields on time scales of order $\frac{1}{Q_S}\ln^2(\frac{C}{\alpha_S})$. The transverse $E$ and $B$ fields may cause large angle deflections of colored particles leading to $p_\perp$ broadening and energy loss of jets--recent numerical simulations by the Frankfurt group appear to confirm this picture~\cite{Frankfurt}. These interactions of colored high momentum particle like modes with the soft coherent classical field modes may also generate a small ``anomalous viscosity" whose effects on transport in the Glasma may
mask a larger kinetic viscosity~\cite{Duke}. The same underlying physics may cause ``turbulent isotropization" by rapidly transferring momentum from soft ``infrared" longitudinal modes to ultraviolet modes~\cite{turbulence}. Finally, albeit the LO result demonstrated that one could have non-trivial Chern-Simons charge in heavy ion collisions, the boost invariance of CYM equations disallows sphaleron transitions that permit large changes in the Chern-Simons number~\cite{KharzeevKV}. With rapidity dependent quantum fluctuations, sphaleron transitions can go. These may have important consequences--in particular $P$ and $CP$ odd metastable transitions that cause a novel ``Chiral Magnetic Effect"~\cite{Warringa} in heavy ion collisions.
Numerical CYM results for Chern-Simons charge and (square root) exponential growth of instabilities are shown in fig.~\ref{fig:LO}.

\subsection{QCD Factorization and the Glasma instability}
\label{sect:four}
\begin{figure}[htb]
\hfill
\includegraphics[width=0.5\textwidth]{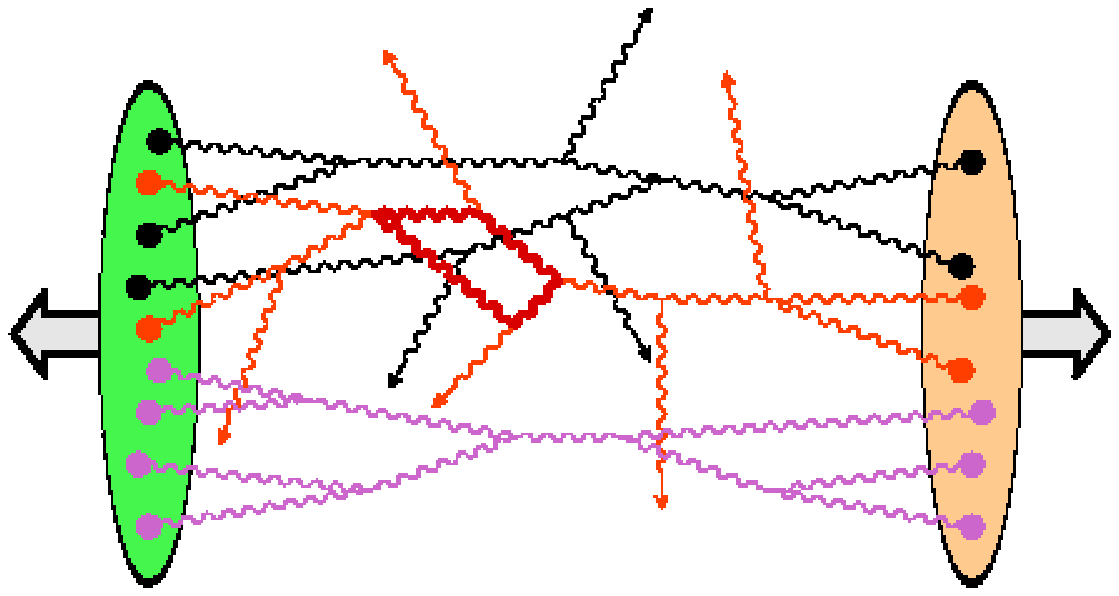}
\hfill
\includegraphics[width=0.2\textwidth]{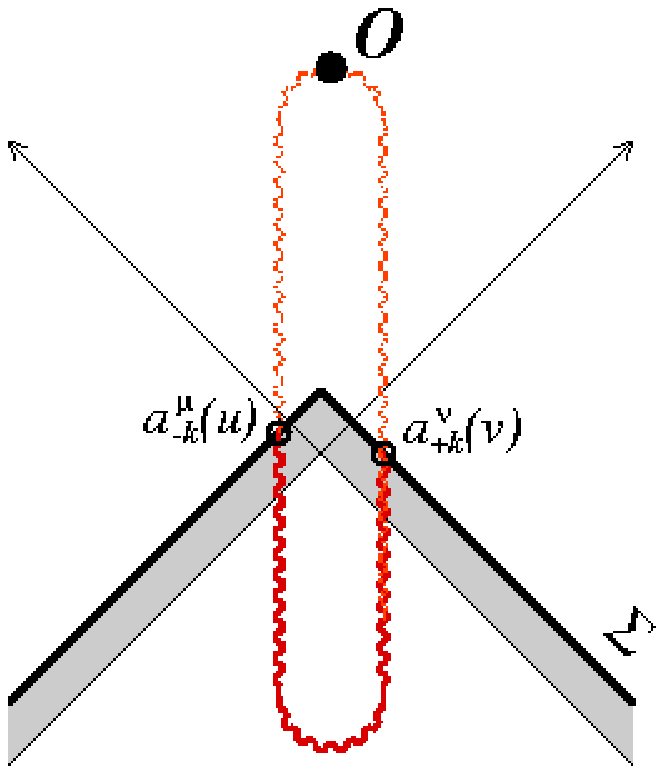}
\hfill
\caption{Left: Illustration of particle production in a field theory with strong time dependent sources. Right: Quantum fluctuations, where the
nuclei talk to each other before the collision, are suppressed; this is responsible for high energy factorization of inclusive gluon operators ${\cal O}$ in A+A collisions~\cite{GelisLV1,GelisLV2}.}
\label{fig:fact}
\end{figure}

The discussion at the end of the last subsection strongly suggests that next-to-leading order (NLO) quantum fluctuations in the Glasma, while parametrically suppressed, may alter our understanding of heavy ion collisions in a fundamental way. To cosmologists, this will not come as a surprise--quantum fluctuations play a central role there as well. In recent papers, it was shown for a scalar theory that moments of the multiplicity distribution at NLO in A+A collisions could be computed as an initial value problem with retarded boundary conditions~\cite{Gelis}; this framework has now been extended to QCD~\cite{GelisLV1,GelisLV2}. In QCD, the problem is subtle because some quantum fluctuations occur in the nuclear wavefunctions and are responsible for how the wavefunctions evolve with energy; others contribute to particle production at NLO. Fig.~\ref{fig:fact} illustrates particle production in fields theories with
strong sources and the non-factorizable quantum fluctuations  that are suppressed in the leading log framework.

In writing eq.~\ref{eq:fact-formula}, its scope of validity was not specified.  A factorization theorem organizing quantum fluctuations shows that all order leading logarithmic contributions to an inclusive gluon operator ${\cal O}$ in the Glasma can be expressed as~\cite{GelisLV1,GelisLV2}
%%%%%%%%%%%%%%%%%%%%%%%%%%%%
\begin{equation}
 \langle {\cal O}\rangle_{_{\rm LLog}}
 = \int [D{\wt{\rho}}_1] [D{\wt{\rho}}_2]\;
W_{Y_{\rm beam}-Y}[\wt{\rho}_1]\, W_{Y_{\rm beam}+Y}[\wt{\rho}_2]\;
 {\cal O}_{_{\rm LO}}\left[\wt{\rho}_1,\wt{\rho}_2\right] \; ,
\label{eq:fact-formula1}
\end{equation}
%%%%%%%%%%%%%%%%%%%%%%%%%%%%%5
where ${\cal O}_{_{\rm LO}}$ is the same operator evaluated at LO by solving classical Yang--Mills equations and 
$W_{Y_{\rm beam}\mp Y}[\wt{\rho}_{1,2}]$ 
are the weight functionals that obey the JIMWLK Hamiltonians describing the rapidity evolution of the projectile and target wavefunctions respectively. This theorem is valid\footnote{Recent work suggests that this factorization theorem is valid even when the rapidity restriction is relaxed.} if the rapidity interval corresponding to the production of the final state, $\Delta Y \leq \frac{1}{\alpha_S}$. The $W$'s are analogous to the parton distribution functions in collinear factorization; determined non-perturbatively at some initial scale $Y_0$, their evolution with $Y$ is given by the JIMWLK Hamiltonian.

This factorization theorem is a necessary first step before a full NLO computation of gluon production in the
Glasma. Eq.~(\ref{eq:fact-formula1}) includes only the NLO terms that are enhanced by a large logarithm of $1/x_{1,2}$, while the complete NLO calculation would also include non enhanced terms. These would be
of the same order in $\alpha_S$ as the production of quark-antiquark pairs \cite{GelisKL} from the classical field. To be really useful, this complete NLO calculation likely has to be promoted to a Next-to-Leading Log (NLL) result by resumming all the terms in $\alpha_S(\alpha_S\ln(1/x_{1,2}))^n$. Now that evolution equations
in the dense regime are becoming available at NLO, work in this direction is a promising prospect.
%%%%%%%%%%%%%%%%%%%%%%%%%%%%%%%%
%%%%%%%%%%%%%%%%%%%%%%%%%%%%%%%%%
In addressing the role of instabilities at NLO, note that small field fluctuations fall into three classes:
i) Zero modes ($p_\eta=0$) that generate the leading logs  resummed in eq.~\ref{eq:fact-formula1}; the coefficients of the leading logs do not depend on $x^\pm$. ii) Zero modes that do not contribute at leading log because they are less singular than the leading log contributions. These become relevant in resumming NLL corrections to the factorization result. Because they are zero modes, they do not trigger plasma instabilities. iii) Non zero modes ($p_\eta\not=0$) that  do not contribute  large logarithms of $1/x_{1,2}$, but grow exponentially  as $\exp(\sqrt{Q_S\tau})$.  While these boost non-invariant terms are suppressed by $\alpha_S$, they are enhanced by exponentials of the proper time after the collision. These leading temporal divergences can be resummed  and the expression for inclusive gluon operators in the Glasma revised to read
\begin{eqnarray}
 &&
\left<{\cal O}\right>_{\rm LLog+LInst}
 =
 \int [D\wt{\rho}_1] [D\wt{\rho}_2]\;
 W_{Y_{\rm beam}-Y}[\wt{\rho}_1]\, W_{Y_{\rm beam}+Y}[\wt{\rho}_2]
 \nonumber\\
 &&\qquad\qquad\qquad\times
 \int\big[Da(\vec u)\big]\;\widetilde{Z}[a(\vec u)]\;
 {\cal O}_{_{\rm LO}}[\wt{\cal A}^+_1+a,\wt{\cal A}^-_2+a]
\label{eq:final}
\end{eqnarray}
where ${\wt{\cal A}}_1^+(x)=-\frac{1}{\partial_\perp^2}\,{\wt{\rho}}_1(x_\perp,x^-)$ and ${\wt{\cal A}}_2^-(x)=-\frac{1}{\partial_\perp^2}\,{\wt{\rho}}_2(x_\perp,x^+)$. The effect of the resummation of instabilities is therefore  to add fluctuations to the
initial conditions of the classical field, with a distribution that depends on the outcome of the resummation. This spectrum
$\widetilde{Z}[a({\vec u})]$ 
is the final incomplete step in determining all the leading singular contributions to particle production in the Glasma-see however Ref.\cite{FukusGM1}. The stress-energy tensor $T^{\mu\nu}$ can then be determined {\it ab initio} and matched smoothly to kinetic theory or hydrodynamics at late times.

\subsection{Two particle correlations in the Glasma and the Ridge}

Striking ``ridge'' events were revealed in studies of the near
side spectrum of correlated pairs of hadrons at RHIC~\cite{Ridge-expt}. The spectrum of correlated pairs on
the near side of the STAR detector extends across the entire detector acceptance in
pseudo-rapidity of order $\Delta \eta\sim 2$ units but is strongly
collimated for azimuthal angles $\Delta \phi$.  Preliminary analysis
of measurements by the PHENIX and PHOBOS collaborations corroborate the STAR results.  In the latter
case, the ridge is observed to span the PHOBOS acceptance in pseudo-rapidity of $\Delta \eta \sim 4$ units.  

\begin{figure}[htb]
\hfill
\includegraphics[width=0.25\textwidth]{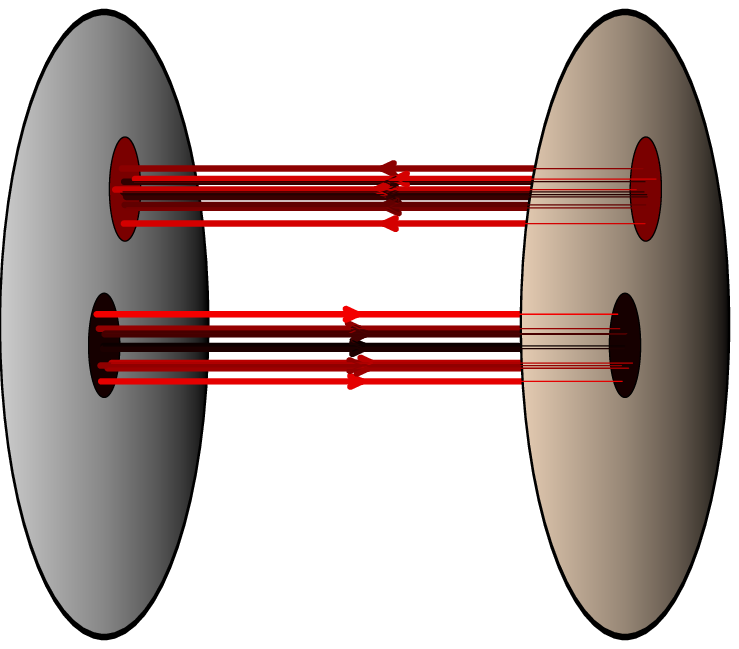}
\hfill
\includegraphics[width=0.45\textwidth]{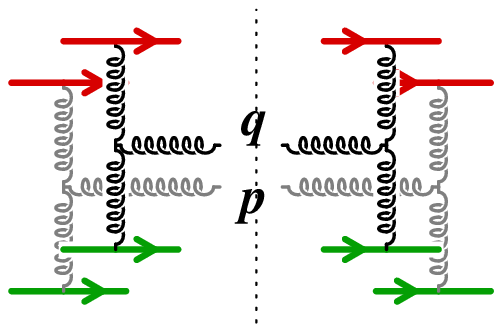}
\hfill
\caption{Left:Glasma flux tubes of transverse size $\frac{1}{Q_S} < \frac{1}{\Lambda_{\rm QCD}}$. The field lines correspond to parallel $E$ and
$B$ fields, which carry topological charge. Right: The leading two particle contribution. Superficially disconnected, they are connected by
averaging over the large x color sources. Systematic power counting shows these graphs dominate over usual ``pQCD" graphs at high energies.}
\label{fig:Ridge}
\end{figure}

Causality dictates (in strong analogy to CMB superhorizon fluctuations) that long range rapidity correlations causing the ridge must have occurred at proper times $\tau\le \tau_{\rm freeze\ out}\;e^{-\frac{1}{2}|y_{_A}-y_{_B}|}$,
where $y_{_A}$ and $y_{_B}$ are the rapidities of the correlated particles. If the ridge span in psuedo-rapidity is large, these correlations must have originated in the Glasma. The PHOBOS data suggest correlations at times as early as a fermi. As noted previously, particles  produced from Glasma
flux tubes are boost invariant. See fig.~\ref{fig:Ridge}. The correlated two gluon production in the Glasma flux tube is independent of rapidity~\cite{DumitruGMV}--thereby allowing long range correlations. Remarkably, the result shown in fig.~\ref{fig:Ridge} (right), that ``classical" disconnected graphs give the leading contribution to two particle correlations, holds true even when one includes leading logarithmic contributions to all orders in perturbation theory~\cite{GelisLV2}. When the separation in rapidity between pairs $\Delta y \gg 1/\alpha_S$, particle emission between the triggered pairs will modify this result.  
This rapidity dependence can be computed and tested at the LHC.
 
Ours is the only dynamical model with this feature-for other models, see Ref.\cite{Hwa}.  The particles produced in a flux tube are isotropic locally in the rest frame but are collimated in azimuthal angle when boosted by transverse flow~\cite{Shuryak-Voloshin}. Combining our dynamical calculation of two particle correlations with a simple ``blast wave" model of transverse flow, we obtained reasonable agreement with 200 GeV STAR data on the amplitude of the correlated two particle spectrum relative to the number of binary collisions per participant pair. 
A more sophisticated recent treatment of the flow of Glasma flux tubes shows excellent agreement of the amplitude of the ridge with centrality and energy, 
and likewise of the angular width of the ridge with centrality and energy~\cite{GavinMM}. Three particle correlations can further help distinguish the 
Glasma explanation for the ridge from other mechanisms.

\section*{Acknowledgments}
 R.V's work on this manuscript was supported by DOE Contract No.~\#DE-AC02-98CH10886. R.V. would like to thank Sourav Sarkar, Helmut Satz and Bikash Sinha or their kind invitation to lecture at the QGP winter school  in Jaipur and for the excellent atmosphere and hospitality at the school. D. B and R.V. would also like to thank the Center for Theoretical Sciences (CTS) of the Tata Institute for Fundamental Research (TIFR) for their support of the
 program and school on ``Initial Conditions in Heavy Ion Collisions" at Dona Paula, Goa, where some of this work was performed.

%%%%%%%%%%%%%%%%%%%%%%%% referenc.tex %%%%%%%%%%%%%%%%%%%%%%%%%%%%%%
% sample references
% "physics"
%
% Use this file as a template for your own input.
%
%%%%%%%%%%%%%%%%%%%%%%%% Springer-Verlag %%%%%%%%%%%%%%%%%%%%%%%%%%

%
% BibTeX users please use
% \bibliographystyle{}
% \bibliography{}

\printindex
\end{document}